\documentclass[prb,twocolumn,showpacs,superscriptaddress]{revtex4-1}
\usepackage{ae}
\usepackage{lipsum}
\usepackage[table]{xcolor}
\usepackage{graphicx}
\usepackage{amssymb}
\usepackage{amsmath}
\usepackage{amsthm}
\usepackage{color}
\usepackage{sidecap}
\usepackage{ulem}
\usepackage{wasysym}
\usepackage{pifont}
\usepackage{blindtext}
\usepackage[caption=false]{subfig}
\usepackage{array}

\newcommand{\up}{\ensuremath{\uparrow}}
\newcommand{\dn}{\ensuremath{\downarrow}}
\newcommand{\Nch}{\ensuremath{N_{\rm ch}}}

\newcommand{\tp}{\ensuremath{t_\perp}}
\newcommand{\kf}{\ensuremath{k_F}}
\newcommand{\Cdag}[3]{\ensuremath{\hat{c}^\dagger_{#1,#2,#3}}}
\newcommand{\C}[3]{\ensuremath{\hat{c}_{#1,#2,#3}}}
\newcommand{\N}[3]{\ensuremath{\hat{n}_{#1,#2,#3}}}
\newcommand{\Ntot}[2]{\ensuremath{\hat{n}_{#1,#2}}}
\newcommand{\ds}{\ensuremath{\Delta_s}}
\newcommand{\dsl}[1]{\ensuremath{\Delta_s[#1]}}
\newcommand{\Ds}{\ensuremath{\overline{\Delta}_s}}
\newcommand{\Dsl}[1]{\ensuremath{\overline{\Delta}_s[#1]}}
\newcommand{\E}[1]{\ensuremath{E_{\rm GS}^{#1}}}

\newcommand{\epsmax}{\ensuremath{\epsilon_{\rm max}}}
\newcommand{\Ltoinf}{\ensuremath{L\rightarrow\infty}}

\begin{document}

\title{Understanding repulsively mediated superconductivity of correlated electrons via massively parallel DMRG}

\author{A.~Kantian}
\affiliation{Department of Physics and Astronomy, Uppsala University, Box 516, S-751 20 Uppsala, Sweden}
\author{M.~Dolfi}
%\affiliation{IBM Research, Zurich, Switzerland, dol@zurich.ibm.com}
%\altaffiliation[Most of the contribution done while at ]{Theoretische Physik, ETH Zurich, 8093 Zurich, Switzerland}
\altaffiliation{Most of this authors contribution took place during his affiliation with ETH Z\"urich}
\affiliation{Theoretische Physik, ETH Zurich, 8093 Zurich, Switzerland}
\affiliation{IBM Research, Zurich, Switzerland, dol@zurich.ibm.com}
\author{M.~Troyer}
\affiliation{Theoretische Physik, ETH Zurich, 8093 Zurich, Switzerland}
\author{T.~Giamarchi}
\affiliation{DQMP, University of Geneva, 24 Quai Ernest-Ansermet, CH-1211 Geneva, Switzerland}

%%%%%%%%%%%%%%%%%%%%%%%%%%%%%%%%%%%%%%%%%%%%%%%%%%
%%%%%%%%%%%%%%%%%%%%%%%%%%%%%%%%%%%%%%%%%%%%%%%%%%

\begin{abstract}
The so-called minimal models of unconventional
superconductivity are lattice models of interacting electrons derived from materials in which electron pairing arises from purely repulsive interactions.
Showing unambiguously that a minimal model actually can have a superconducting ground state remains a challenge at nonperturbative interactions. We make a significant step in this direction by computing
ground states of the 2D \mbox{U-V} Hubbard model - the minimal model of the quasi-1D superconductors - by parallelized DMRG, which allows for systematic control of any bias and that is sign-problem-free. Using distributed-memory supercomputers and leveraging the advantages of the \mbox{U-V} model, we can treat unprecedented sizes of 2D strips and extrapolate their spin gap both to zero approximation error and the thermodynamic limit. 
Our results for the spin gap are shown to be compatible with a spin excitation spectrum that is either fully gapped or has zeros only in discrete points, and conversely that a Fermi liquid or magnetically ordered ground state is incompatible with them. Coupled with the enhancement to short-range correlations that we find exclusively in the $d_{xy}$ pairing-channel, this allows us to build an indirect case for the ground state of this model having superconducting order in the full 2D limit, and ruling out the other main possible phases,  magnetic orders and Fermi liquids.
\end{abstract}
		
\date{\today}
\maketitle

%%%%%%%%%%%%%%%%%%%%%%%%%%%%%%%%%%%%%%%%%%%%%%%%%%
%%%%%%%%%%%%%%%%%%%%%%%%%%%%%%%%%%%%%%%%%%%%%%%%%%

\section{Introduction}\label{sec:intro}

Understanding the impact of inter-particle repulsions in fermionic quantum systems remains one of the most challenging problems in physics.
As was shown by Landau, electron-electron repulsions in a solid may often be treated by redefining excitations as quasiparticles. The resulting physics is thus essentially of free particles (the so-called Landau quasiparticles), with interactions mostly leading to
a renormalization of some physical parameters such as the mass. This Fermi liquid (FL) approach has seen very broad success and allowed to understand the effects caused by perturbations on top of the Fermi liquid, such as instabilities towards e.g. a Bardeen-Cooper-Schrieffer (BCS) superconducting (SC) state, a magnetically ordered (MO) state or a charge density wave (CDW) state~\cite{BookZiman1972}.

There are however situations where the FL approach breaks down, leading to so-called ``non-Fermi liquid'' behavior. At high spatial dimensionality, this can happen if the interactions are particularly strong, or if the filling of the systems is commensurate with the lattice, with e.g. one particle per
site leading to a Mott (MI) state. Reduced dimensionality of the system can further enhance the effect of interactions. In one dimension in particular, the interactions always have drastic effects and lead to physical properties very different from the ones of a Fermi liquid. The FL is replaced by another set of universal features, called the Tomonaga-Luttinger liquid (TLL), where the elementary excitations are the collective excitations of charge and spin in the system.
Such systems are critical and at zero temperature $T=0$ possess various competing quasi-long range orders, ranging from antiferromagnetism to superconductivity~\cite{BookGiamarchi2003}.

Non-FLs have thus been prime candidates to search for novel physical phases, in particular unconventionally superconducting (USC) phases mediated by excitations other than the electron-phonon one, of which the high-temperature (high-$T_c$) superconductors~\cite{BookAnderson1997,Orenstein2000}. are just one example. The observed proximity between USC and MO phases has made fluctuations around the MO state prominent candidates for such pairing mechanisms. On the experimental side, many systems ranging from heavy fermions~\cite{Varma1985,Si2010}, organic superconductors~\cite{Bourbonnais2007,Jerome2012}, high-$T_c$ superconductors and pnictides~\cite{Damascelli2003,Fischer2007,Scalapino2012a}  show USC phases.
Yet, obtaining any widely accepted theory on the nature of such exotic pairing for any of these material-groups has proven to be difficult; even the simplest physical models, the so-called \textit{minimal models}, which are abstracted from the materials' full physical structure are fundamentally hard to solve: the 2D \mbox{U-V} model at half filling (organics), and the weakly doped 2D Hubbard model (cuprates). So far, any solution of these minimal models has almost always necessitated additional technical approximations that introduce errors of unknown magnitude. It is thus difficult to determine whether phases predicted for such models in the framework of a given approximation scheme are truly present or an artefact of the
approximation, or whether to explain the experiments, different, more extended models are required, such as e.g. in the three-band models of the high-$T_c$ cuprates~\cite{Varma1991}, or in those approaches that investigate the role of phonons in USC materials~\cite{BookSalje1996}.

Even when moving to the more tractable regime of weak repulsion, SC and MO instabilities interfere at all orders of the perturbative renormalization group (RG) treatments applied so far~\cite{Schulz1987,Furukawa1998,Lauchli2004}, necessitating complex numerical methods~\cite{Deng2015} or sophisticated, and still approximate, functional RG procedures~\cite{Nickel2005,Nickel2006,Bourbonnais2011,Sedeki2012,Metzner2012a} .
More uncertain yet is the situation in the most important regime, actually relevant to the models' underlying materials: when repulsion becomes large and competes with kinetic energy. For this regime, numerical methods are a route of choice. However, in two or more spatial dimensions these too are faced with steep challenges. Quantum Monte Carlo (QMC) treatments suffer from the so-called sign problem for fermions leading to an exponential degradation of the signal to noise ratio as e.g. the temperature is lowered~\cite{LeBlanc2015}. And while innovative techniques like diagrammatic QMC have delivered intriguing insights into USC pairing for the 2D Hubbard model, these are still constrained to intermediate interactions at most and doping very far from unity~\cite{Deng2015}.

Our numerical approach here is different, and rests on two observations: (1) The fact that for the Density Matrix Renormalization Group (DMRG) technique, and other tensor-network methods such as PEPS, any possible bias can be controlled against via extrapolation, and that these methods are free of the sign problem~\cite{Schollwock2011,Stoudenmire2012}. (2) That the one minimal 2D model in which a USC ground state from strong electron repulsions can unambiguously be shown to exist, is comprised of quasi-1D electrons (doped \mbox{2-leg} Hubbard ladders) weakly coupled in parallel~\cite{Carlson2000,Karakonstantakis2011}. This minimal model however, as yet, does not correspond to any extant material. 

These two observations have led us to conclude that the minimal 2D \mbox{U-V} model of the organic superconductors - which has an analogous structure, of many 1D systems coupled weakly in parallel to each other - offers an unique opportunity for insight on the strongly interacting ground state of a candidate model for USC pairing. This would be done by swapping out the perturbative theory used on the Hubbard-ladders in the low-energy field-theory limit, which would not be applicable to the \mbox{U-V} model setting, with quantitatively reliable DMRG. 

Yet, using DMRG in this particular setting comes with issues to resolve first. While DMRG delivers highly accurate results both for the statics and the dynamical correlations in 1D systems while using only modest computational resources, the 2D setting of the \mbox{U-V} model is different. On such lattices, DMRG is known tho require resources that increase quasi-exponentially with lattice-width when trying to maintain any pre-set accuracy~\cite{Stoudenmire2012}.
Tractable sizes for systems of itinerant electrons so far usually have been on the order of about 100 sites, depending on the specific problem~\cite{White2009,Scalapino2012,Liu2012,LeBlanc2015,Ehlers2017,Zheng2017} - pushing towards around 200 sites, while done, is generally challenging and goes along with a decrease in accuracy. As the sum of the evidence indicates that states with SC order are in close energetic competition with MO ones, even changes in geometry (i.e. boundary effects) could easily result in different orders winning out for lattices of such size. As a result, no unbiased approach has so far been able to show unambiguously that a ground state of either of the two minimal models may exhibit USC order when repulsion is non-perturbatively large.

In the present work, we take a significant step in that direction. We build on our development of the numerical parallelized density matrix renormalization group (pDMRG), a distributed-memory version of one of the standard DMRG formulations~\cite{Schollwock2011} capable of 
exploiting modern supercomputer architectures.
With this method we investigate the properties of the 2D \mbox{U-V} Hamiltonian at half filling of the bands. We consider an array of parallel chains of strongly correlated 1D electrons, each described by such a U-V Hamiltonian and weakly coupled to each other by transverse tunnelling.
In addition to being the model Hamiltonian for organic superconductors~\cite{Giamarchi2004}, and thus potentially containing the unconventionally superconducting phase observed in this system,
this Hamiltonian has two critical advantages for numerical study compared to the doped 2D Hubbard model~\cite{LeBlanc2015}: (1) it is trivial to maintain fixed ratios of electrons to lattice sites and thus to extrapolate to infinite system size here, since no doping is required. The pDMRG then allows doing this with the accuracy required to extrapolate energies of large systems, such as to enable the reliable calculation of energy differences, as appearing in e.g. the most important observable we use, the \textit{spin gap}. Underlying this accuracy is that pDMRG can exploit the strong anisotropy of electron tunneling in the 2D \mbox{U-V} model by spreading out a single ground state calculation across many supercomputer nodes. This anisotropy ties directly into the U-V models' second advantage: (2) the strip or cylinder geometry,  that DMRG is especially good at handling intrinsically, suits the U-V model naturally if aligned with the strong-tunneling direction, as correlations across the strips' width will be naturally weak due to the small perpendicular tunneling, which is not the case for the doped 2D Hubbard model.

We show that in the thermodynamic limit this systems' averaged spin gap $\overline{\Delta}_s$ is incompatible with either a FL or an MO ground state, for strips of finite width. The behaviour of $\overline{\Delta}_s$ as $L\rightarrow\infty$ appears most compatible with $\overline{\Delta}_s>0$ at infinite length, and being either non-decaying or even increasing with the strips' width, i.e. pointing towards a fully gapped spin excitation spectrum. But we also consider the alternate possibility that $\overline{\Delta}_s$ might ultimately scale to zero beyond the system sizes we can access. However, we show this alternative is still at most compatible with a spin excitation spectrum with isolated zero-energy points, i.e. in line with weak-coupling theories of USC systems. But either scenario is incompatible with either antiferromagnetic order or Fermi liquid-like ground state.
Together with the strong enhancement exclusive to the $d_{xy}$-channel of correlations that we find with increasing strip width, our results make the case for SC singlet pairing with $d_{xy}$-order being the dominant component in the ground state of the \mbox{U-V} model at half filling when repulsion is competitive with kinetic energy.

The structure of the paper is as follows: in section~\ref{sec:over} we outline the \mbox{U-V} model and our approach to analysing the system and choosing the model parameters; in section~\ref{sec:method} we describe the key features of pDMRG relevant to this work; in section~\ref{sec:scale} we detail our scaling procedure; in section~\ref{sec:res} we analyse the behaviour of the spin gap and the correlation functions as the width of the 2D \mbox{U-V} strips is increased, making the case for a SC ground state with $d_{xy}$-order and against any MO or FL competitor-order; in section~\ref{sec:disc} we discuss these results and how they connect to experiment; in Appendix~\ref{app:scale} we provide details  for the used scaling procedure; finally, in Appendix~\ref{app:dmrg_tech} we provide a broader technical overview on pDMRG.

%BEGIN FIGURE%
\begin{figure}[h]
\centering
\includegraphics[width=1\columnwidth]{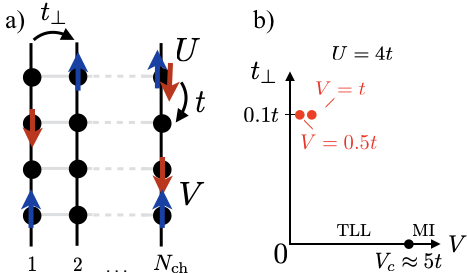}
\caption{
\textbf{(a)} Overview of the 2D \mbox{U-V} model. Conduction electrons tunnel along effective 1D chains with amplitude $t$, where electrons are subject to strong on-site repulsion $U$ and a weaker nearest-neighbour repulsion $V$. Inter-chain tunneling $\tp$ is much weaker than $t$.
\textbf{(b)} Schematic parameter space of the 2D \mbox{U-V} model at half filling, for fixed $U=4t$. The only limit in which all possible ground states
and the transition between them is understood is the 1D limit, i.e. at $\tp=0$, when the chains decouple, and are either in a TLL or MI phase, depending
on whether $V<V_c$ or $V>V_c$. The red dots show the parameters at which we work in this present article.}
\label{fig:over}
\end{figure}
%END FIGURE%
%%%%%%%%%%%%%%%%%%%%%%%%%%%%%%%%%%%%%%%%%%%%%%%%%%%
%%%%%%%%%%%%%%%%%%%%%%%%%%%%%%%%%%%%%%%%%%%%%%%%%%

\section{Model and physical observables}\label{sec:over}

As depicted in Fig.~\ref{fig:over}a, the \mbox{U-V} model in 2D is formed by a parallel array of $\Nch$ 1D chains of lattice electrons, each of length $L$, with inter-chain tunneling.
Its lattice-Hamiltonian is

\begin{align}\label{eq:ham}
\hat{H} & = -t \sum_{i=1}^{L-1}\sum_{n=1}^{\Nch} \sum_{\sigma=\up,\dn} \left(\Cdag{i}{n}{\sigma} \C{i+1}{n}{\sigma} +\mbox{h.c}\right) \nonumber \\
 & -\tp \sum_{i=1}^{L}\sum_{n=1}^{\Nch-1} \sum_{\sigma=\up,\dn} \left(\Cdag{i}{n}{\sigma} \C{i}{n+1}{\sigma} + \mbox{h.c}\right) \nonumber \\
 & +U \sum_{i=1}^{L}\sum_{n=1}^{\Nch}  \N{i}{n}{\up} \N{i}{n}{\dn}  + V \sum_{i=1}^{L-1}\sum_{n=1}^{\Nch}  \Ntot{i}{n}\Ntot{i+1}{n}
\end{align}

Here, $\Cdag{i}{n}{\sigma}$ ($\C{i}{n}{\sigma}$) denotes the electron creation (annihilation) operator on site $i$ of chain $n$ of spin $\sigma=\up,\dn$, and $\N{i}{n}{\sigma}:=\Cdag{i}{n}{\sigma}\C{i}{n}{\sigma}$, $\Ntot{i}{n}:=\N{i}{n}{\up}+\N{i}{n}{\dn}$ denote local electron density for spin $\sigma$ and total electron density operators respectively. The tunneling amplitude along the chains is given by $t$, while in-between adjacent chains it is given by $\tp$, and generally $t\gg\tp$. We point out that the $\tp$-term above obeys open boundary conditions, just as the $t$-term does. We have made this choice because obtaining ground state energies accurate enough to compute reliable differences between them is an indispensable requirement to extrapolate this systems spin gaps in the limit $\Ltoinf$. Cylindrical 2D-lattices however are well known to immediately suffer from lower accuracy, and are thus avoided here. Finally, the onsite Coulomb repulsion is given by $U$, while intrachain nearest-neighbour repulsion is $V$, with $V<U/2$. The entire system is at half-filling, i.e. $N_\up = N_\dn = L\Nch/4$, and thus $\kf^\uparrow=\kf^\downarrow=\kf=\pi/4a$, where $a$ is the lattice spacing. 

This minimal model originally arose from the study of the organic Bechgaard and Fabre salts (`the organics'), the first materials discovered to support an USC phase~\cite{Jerome1989,Bourbonnais2007}. These compounds show a very rich phase diagram, and especially a change in effective dimensionality with temperature - 1D TLL-like at high temperature, then 2D-like and finally 3D-like at near zero - as quantum coherence can increasingly be established along the three orthogonal and successively weaker directions for electron-tunneling. Their most striking feature is the phase transition from a magnetically ordered to a unconventionally superconducting phase at low temperature, as the proximity, and thus the inter-chain tunneling $\tp$, between the strong-coupling chains of the \mbox{U-V} model is increased~\footnote{Experimentally, this distance is controlled either by applying external pressure to the sample, or chemically (choice of anions between the cationic stacks of molecules that make up the 1D chains)}.
To understand this competition, and whether the minimal \mbox{U-V} model~(\ref{eq:ham}) of the organics can actually capture it at least in a weak-interaction scenario (which does not correspond to the actual materials), perturbative RG has been repeatedly applied~\cite{Nickel2005,Nickel2006,Bourbonnais2011,Sedeki2012}. While beset with the same technical limitations this approach has for the doped 2D Hubbard-model of the cuprates (c.f. Sec.~\ref{sec:intro}), it does suggest that the minimal 2D \mbox{U-V} model, might indeed support a USC phase transition from a MO phase when extended with e.g. a next-to-nearest neighbour tunneling perpendicular to the chains, and that pairing would be in the $d_{x^2-y^2}$-channel (gapless superconductivity would be compatible with experiments on the organics~\cite{Shinagawa2007,Yonezawa2012}). However, so far there has been no complementary method to either validate these results, and especially no numerics that could deal with the \mbox{U-V} model~(\ref{eq:ham}) for the experimentally relevant case of strong interactions.

In order to be able to address the case $\tp\neq 0$, $\Nch>2$ for this Hamiltonian with reliable numerics, we have developed pDMRG to exploit the distributed-memory architectures of modern supercomputers. Based on the tensor-network formulation widely used in modern DMRG, the code has major advantages over other quantitative numerical approaches for this problem: it inherits the lack of sign-problem and the ability to use extrapolation in the known error to filter out possible implicit bias towards any particular order (which is inherent e.g. to any wavefunction ansatz of variational quantum Monte Carlo) from conventional DMRG. And further, as described in Sec.~\ref{sec:method} and Appendix~\ref{app:dmrg_tech}, the parallelized nature of pDMRG makes the necessary ground state calculations tractable in the first place, by distributing the many non-local tunneling terms of the model across different nodes of the supercomputer. As laid out in Sec.~\ref{sec:method}, the large number of such terms becomes unavoidable when keeping the amount of bipartite entanglement at manageable levels, which is performance-critical for any 2D model. 

As detailed in Sec.~\ref{subsec:obs}, the spin gap is the most important observable for our study of this system, as its scaling as $\Ltoinf$ allows the crucial elimination of possibilities for ground state ordering in the thermodynamic limit that has so far proven elusive for any minimal model of USC systems. This is because ground states with either MO order or in a conventional FL state will show a linear vanishing of any spin-gap. In contrast, standard analytical theories of USC systems combine weak-coupling RG, for identifying the leading instability and its symmetry, with mean-field descriptions of the ordered phase. For these theories, the unavoidable zero-nodes of the order parameter result in sublinear scaling to zero for a spin gap~\cite{Sigrist1991,Hasegawa1996,Moriya2003,Shinagawa2007,Chubukov2008,Vladimirov2011}, while experiments on the actually strongly coupled regime in USC materials have found finite spin gaps (c.f. Sec.~\ref{sec:disc}). It is therefore vital that the known maximal value of the approximation error of the DMRG technique allows us to extrapolate the computed spin gaps to zero truncation error for any given lattice size, as described in~\ref{subsec:truncerr}. 

To outline our strategy of analysis, we now discuss in Sec.~\ref{subsec:obs} the concrete ground state observables we considered, providing the roadmap for ruling out the main competitor states to any superconducting ground states, the FL and the MO states. In Sec.~\ref{subsec:parms} we then discuss our choice of parameters for Hamiltonian~(\ref{eq:ham}), and why these should reduce the competition with charge-ordered ground states.

%%%%%%%%%%%%%%%%%%%%%%%%%%%%%%%%%%%%%%%%%%%%%%%%%%

\subsection{Observables}\label{subsec:obs}

One of our key priorities is to establish whether any set of parameters for Hamiltonian~(\ref{eq:ham}) could or could not result in one of the main competitor orders to a USC ground state in the 2D limit, a MO or FL state, being realized.  One quantity that would allow making these distinctions is the
spin susceptibility at zero temperature
\begin{widetext}
\begin{equation} \label{eq:spinsuc}
 \chi^{-1}_s(L,\Nch) = L \frac{\E{L,\Nch}(S_z = s) + \E{L,\Nch}(S_z = -s) - 2 \E{L,\Nch}(S_z = 0)}{(2 s)^2}
\end{equation}
\end{widetext}
where $\E{L,\Nch}$ denotes the ground state energy of (\ref{eq:ham}) at half-filling, and $S_z=N_\uparrow-N_\downarrow$ is the difference between spin-populations.
Here, $s$ can be any spin number, as long as it is negligible compared to a macroscopic number of spins (proportional to $L$).
Typically, $s$ is chosen so that other quantum numbers in the system can be kept identical between $s$ and $s=0$. The usual
choice is $s=1$, corresponding to the spin flip of a single spin. However, as discussed in subsection \ref{subsec:remosc}, the choice
$s=\Nch$ has a crucial technical advantage for extrapolating to infinite length $L\rightarrow\infty$.

When $L\to\infty$, $\chi_s(L,\Nch)$ becomes the thermodynamic spin susceptibility.
In order to get a finite susceptibility, as expected for a FL or a MO state, one would thus need the
quantity
\begin{eqnarray} \label{eq:endif}
 A(L,\Nch,s) & = & \E{L,\Nch}(S_z = s) + \E{L,\Nch}(S_z = -s) \nonumber \\
 && - 2 \E{L,\Nch}(S_z = 0)
\end{eqnarray}
to scale as $1/L$. Other scalings of this quantity
thus directly signal a non-FL, non-MO behavior for the spin susceptibility. The simplest case is that the \textit{averaged spin gap}
\begin{equation}\label{eq:Ds}
\Dsl{L,\Nch} := \frac{ \E{L,\Nch}(S_z=\Nch) - \E{L,\Nch}(S_z=0)}{\Nch},
\end{equation}
remains finite when $L\to \infty$, where $\Dsl{L,\Nch}$ is the generalization of the standard spin gap
\begin{equation}\label{eq:ds}
\dsl{L,\Nch} := \E{L,\Nch}(S_z=1) - \E{L,\Nch}(S_z=0),
\end{equation}
Then the spin susceptibility would scale as ${(L\Dsl{L\rightarrow\infty,\Nch})^{-1}}$, leading to zero susceptibility in the thermodynamic limit, as expected for fully gapped systems. 
For SC systems with unconventional electron pairing such fully gapped spin-excitations have been actually been found experimentally (see Section~\ref{sec:disc}).

More complex behavior can be expected if (\ref{eq:endif}) has a different scaling with the size of the system such as e.g. $1/L^\alpha$ with $0 < \alpha < 1$. The thermodynamic spin susceptibility remains zero in this case, yet the systems' spin excitations are not fully gapped. In mean-field theories of USC systems all this stems from the line- or point-nodes of the SC gap~\cite{Sigrist1991,Hasegawa1996,Moriya2003,Shinagawa2007,Chubukov2008,Vladimirov2011}. At finite temperatures, the scaling with $L$ would be converted into a non-trivial temperature dependence of the spin susceptibility corresponding to a pseudogap behavior~\cite{James2013,Chen2017a}.

Establishing the $L$-dependence of $\Dsl{L\rightarrow\infty,\Nch}$ through our numerical results therefore allows to differentiate between three possibilities for the ground state of the two dimensional system in the $T=0$ limit: FL/MO, or fully spin gapped USC, or mostly spin-gapped USC  - to the extent that we can extrapolate trends as $\Nch$ is increased.

The use of pDMRG allows meaningful extrapolation of $\Dsl{L\rightarrow\infty,\Nch}$ in $1/L$ for the first time, enabling the calculation of ground states of multiple coupled, long  \mbox{U-V} chains ($\Nch \leq 8$, $L\leq 64$)
with quantitatively accurate numerics (c.f. Sec.~\ref{sec:method})~\footnote{ $\Nch$ larger than presented in this work can be handled by pDMRG. As a single calculation can be spread out across dozens of nodes i.e. thousands of CPU cores, a given problem is less limited by the capacity of the computer as such, but by the available time on the supercomputer.}. As discussed in Sec.~\ref{subsec:spingap}, this allows us to make a strong case that for our choice of Hamiltonian parameters any magnetic order or Fermi liquid behaviour is suppressed in the ground state of the 2D \mbox{U-V} model, due to the way $\Dsl{L,\Nch}$ behaves with $L\rightarrow\infty$, and showing this beviour to be robust as $\Nch$ grows. 

Our point of reference for this are 2D arrays of coupled spin-chains, which demonstrate the opposite scenario, of the spin gap decaying with $\Nch$. If infinitely many 1D Heisenberg-chains were to be coupled in parallel, the result is a gapless 2D Heisenberg model - but when only two such chains are coupled this system exhibits a finite spin gap~\cite{Haldane1983a,Dagotto1996}. Multiple techniques have since shown how these two extremes are bridged: as the number of coupled Heisenberg chains increases, the spin gap remains finite (for an even numbers of chains) but continuously decreases as $\Nch$ grows, scaling to zero in the limit of infinitely many chains~\cite{Schulz1986,Affleck1989}. 

The information gained from spin gaps can be supplemented by studying the change in the short-range behaviour of certain correlation functions
as $\Nch$ is increased. When magnetic orders can be ruled out, the functions of interest here are those of  the two other main competitor groups: SC-ordering on the one hand - with the possibility of pairing in the $d_{x^2-y^2}$-, $d_{xy}$- and extended $s$-wave channels - and CDW-ordering on the other hand. The corresponding correlation functions considered for this are
\begin{widetext}
\begin{eqnarray}
\label{eq:dx2y2corr} d_{x^2-y^2}(r) & := & \left\langle \left( \hat{D}^\dagger_{L/2,\Nch/2,0,1} - \hat{D}^\dagger_{L/2,\Nch/2+1,1,0} \right) \times \left( \hat{D}_{L/2+1+r,\Nch/2,0,1} - \hat{D}_{L/2+1+r,\Nch/2+1,1,0} \right)\right\rangle \\
\label{eq:dxycorr} d_{xy}(r),s(r) & := & \left\langle \left( \hat{D}^\dagger_{L/2+1,\Nch/2,-1,1} \mp \hat{D}^\dagger_{L/2+1,\Nch/2,1,1} \right) \times\left( \hat{D}_{L/2+3+r,\Nch/2,-1,1} \mp \hat{D}_{L/2+3+r,\Nch/2+1,1,1} \right)\right\rangle \\
\label{eq:ddcorr} C(r) & := & \left\langle \hat{n}_{L/2,\Nch/2} \hat{n}_{L/2+r,\Nch/2} \right\rangle - \left\langle \hat{n}_{L/2,\Nch/2} \right\rangle \left\langle\hat{n}_{L/2+r,\Nch/2} \right\rangle,
\end{eqnarray}
\end{widetext}
where $\hat{D}$ denotes the nearest-neighbour spin-singlet operators $\hat{D}_{i,n,j,m}:=\C{i}{n}{\uparrow}\C{i+j}{n+m}{\downarrow}-\C{i}{n}{\downarrow}\C{i+j}{n+m}{\uparrow}$. As illustrated in Fig.~\ref{fig:corrdecay}b, we consider the SC correlation functions
in the two central chains $\Nch/2$, $\Nch/2+1$, and $C(r)$ on chain $\Nch/2$, in each case at distance $r$ from the middle of the chain(s).
Based on these correlation functions, we present further evidence in Sec.~\ref{subsec:corrs} beyond the behaviour of the spin gap that for the \mbox{U-V} model we study here, electrons do in fact pair, with a dominant component in the $d_{xy}$-channel.

%%%%%%%%%%%%%%%%%%%%%%%%%%%%%%%%%%%%%%%%%%%%%%%%%%

\subsection{Choosing model parameters - $U=4t$, $V/t = 0.5,1$}\label{subsec:parms}
%BEGIN FIGURE%
\begin{figure}[h]
\centering
\includegraphics[width=1\columnwidth]{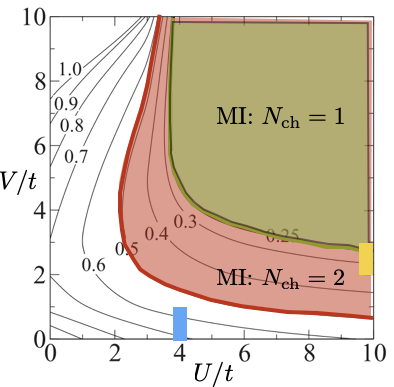}
\caption{
\mbox{U-V} parameter space. Green shaded area: single chain enters MI-regime. Red shaded area: two coupled chains can enter MI-regime~\cite{Kantiana}. To be certain we are outside the regime where this counter-intuitive increase in parameter space for the MI-insulator takes place in the few-chain regime, we work within the blue-shaded area of parameter space. For reference, the yellow shaded area shows the parameter regime realized in the Bechgaard and Fabre salts. Black lines show lines of constant $K_\rho$ for the single \mbox{U-V} chain (Data and figure reprinted from~\cite{Ejima2005}).
}
\label{fig:parms}
\end{figure}
%END FIGURE%

The basic results serving as background to any choice of Hamiltonian parameters are summarized in Fig.~\ref{fig:over}b. In the limit $\tp=0$, the chains decouple and are described by TLL-theory for sufficiently small $V$. As long as $U$ is fixed to any value $\geq 2t$, there exists a value $V_c$ above which a charge gap opens in the system, turning it into a Mott insulator (MI)~\cite{Mila1993,Ejima2005}. The phase boundary at $\tp=0$ is shown as a green line in Fig.~\ref{fig:parms}. Now, in order to maximise the chance of finding SC order, we attempt to target a parameter regime in which the decoupled chains would be in the TLL phase at the same time as the coupled chains ($\tp>0$) would remain in a delocalized regime. Ultimately, only the computational results can indicate whether one succeeds with the latter condition. But given the limited supercomputing time at hand, we try to maximise our chances for this to happen from the outset, as described within the rest of this section.

A look at the minimal \mbox{U-V} model of the organic superconductors (Bechgaard and Fabre salts) is instructive here. Experimental probes of these show that here the 1D chains of the \mbox{U-V} model have a Luttinger-liquid parameter $K_\rho$ (which characterizes interactions and correlations of the charge mode of an isolated chain) in the vicinity of $0.25$~\cite{Giamarchi2004}. When $K_\rho= 0.25$, theory predicts the single chain at half-filling to develop a charge gap and become a MI. In terms of the microscopic parameters, this vicinity to $K_\rho=0.25$ for the single chain would correspond to $U/t=10$, $V/t\approx 2 - 3$, c.f. yellow area in Fig.~\ref{fig:parms}. When the ratio $\tp/t$ is small enough, experiments show the physics of the single chain extending to the whole material, i.e. the system is an MI as well, and magnetically ordered on top of it. Only as $\tp/t$ grows and $\tp$ becomes competitive with the Mott gap of the single chain is superconductivity found in experiments, typically at $\tp/t\approx 0.1$.

It may be tempting to try to straightforwardly replicating this setting for our calculations. However, there has recently been new insight on the two-chain limit by two co-authors of the present work~\cite{Kantiana}, which cautions against such an approach. Combining analytical RG treatment of the bosonized chains with DMRG, we find that finite $\tp$ for two chains \textit{increases} the \mbox{U-V} parameter space in which the system develops a Mott gap (red line in Fig.~\ref{fig:parms}). While early exploratory work indicates that this parameter space might shrink again for $\Nch=4$, it is beyond the scope of this work or of Ref.~\onlinecite{Kantiana} for definite statements how exactly the few-chain regime evolves in this regard.

In the following, we therefore focus on a regime in which these complications of few-chain physics are unlikely to apply, at $U/t=4$ and $V/t=0.5,1$, and with $\tp/t=0.1$ (blue shaded region in Fig.~\ref{fig:parms}). This regime realizes the essential physics present
in the superconductivity of the organics, i.e. a regime where inter-chain tunneling is not effectively suppressed by a single-chain gap. A priori, this regime thus appears to offer the highest chance of finding a superconducting ground state due to repulsively mediated pairing. We have targeted two values of $V/t$ to be certain our results will be representative for a finite area of parameter space.

%%%%%%%%%%%%%%%%%%%%%%%%%%%%%%%%%%%%%%%%%%%%%%%%%%
%%%%%%%%%%%%%%%%%%%%%%%%%%%%%%%%%%%%%%%%%%%%%%%%%%

\section{The Method: parallel DMRG}\label{sec:method}
%BEGIN FIGURE%
\begin{figure}[h]
\centering
\includegraphics[width=1\columnwidth]{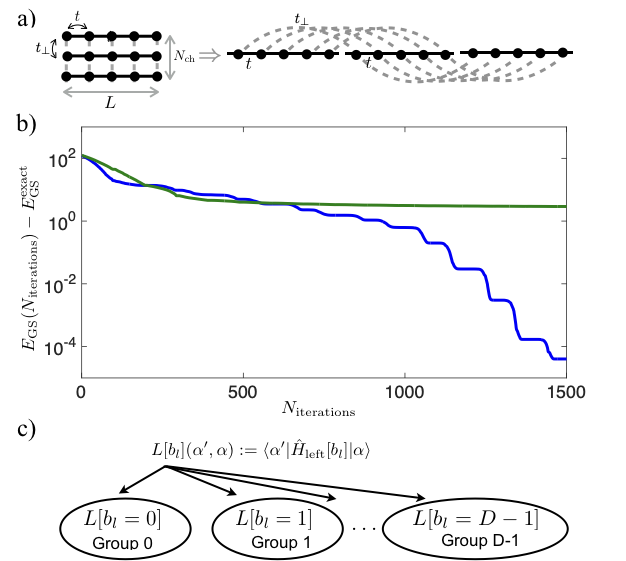}
\caption{
\textbf{(a)} Mapping the physical 2D lattice to the 1D DMRG chain, along the $t$-direction.
Many more long-range tunneling terms appear than would be present when mapping along the $\tp$-direction.
\textbf{(b)} How the two different mappings perform: mapping along $t$ (blue) rapidly converges to the exact solution, while
$t_p$-mapping is trapped in a metastable state, even at $\chi=5000$ and many sweeps.
{\bf (c)} The primary parallelization layer of pDMRG used for the present work, using the left boundary $L$ as an example. Each element $L[b_l]$ of $L$ is dispatched to a MPI-process, and each such MPI-process may control a number of physical nodes. The process then locally handles contractions of $L[b_l]$  with the wavefunction tensors to be optimized, requesting the results of other MPI-processes for different $b_l$ as necessary.
}
\label{fig:pdmrg}
\end{figure}
%END FIGURE%
As stated in Sec.~\ref{sec:over}, exploiting the physical structure of the 2D \mbox{U-V}-model is essential for being able to handle the bipartite entanglement of this system when mapping the physical lattice to the DMRG-chain. Being comprised of many strongly
correlated 1D chains in parallel with weak coupling in-between, it would seem intuitive that performing the mapping along the direction of strong tunneling (c.f. Fig.~\ref{fig:pdmrg}a) delivers lower overall bipartite entanglement in the 1D DMRG chain. This argument is complemented by theoretically estimating the performance of the alternate mapping along the $\tp$-direction that is used conventionally when applying DMRG to what is effectively a long and narrow 2D-strip. The \textit{bond dimension}, denoted as $\chi$ in the following, is the central quantity that controls the accuracy of DMRG. It denotes the number of optimally chosen basis states in which any bipartition of the system is represented. If $\chi_{\rm chain}$ was required to represent a single \mbox{U-V}-chain of a given length with truncation error $\epsilon$, achieving the same accuracy for $\Nch$ chains would require $\chi=\chi_{\rm chain}^{\Nch}$ - even in case $\chi_{\rm chain}=10$, this scaling would be hopeless beyond $\Nch=2$. A simple test comparing the two mappings for a representative system of free fermions, summarized in Fig.~\ref{fig:pdmrg}b, confirms the total disparity between the performance of these mappings. Shown is the energy vs. the number of DMRG iterations for $L=20$, $\Nch=5$, $N_\up=N_\dn=50$, $\tp/t=0.2$, $U=V=0$, $\chi=5000$, once with mapping along $t$ (blue line), once with mapping along $\tp$(green line). The $t$-mapping converges quickly to the exact ED solution, while the $\tp$-mapping flounders even for this small system. 

The price paid for the $t$-direction mapping, necessary to have manageable $\chi$ in the first place, is the very large number of long-range tunneling terms. In the two-site DMRG we use, for every adjacent pair of sites ${(i,i+1)}$ on the DMRG-chain for which the tensors are jointly optimized, the Hamiltonian is represented as a matrix product operator (MPO)~\cite{Schollwock2011}:
\begin{equation}\label{eq:mpo}
\hat{H}=\sum_{b_l,b_c,b_r=1}^D \hat{H}_{\rm left}[b_l] \otimes \hat{h}_{i}[b_l,b_c] \otimes \hat{h}_{i+1}[b_c,b_r] \otimes  \hat{H}_{\rm right}[b_r]
\end{equation}
Here, each sum runs from $1$ to $D$, the MPO bond-dimension, which for the $t$-direction mapping scales as $D\approx 4L$. For the range of system lengths we treat, $L=20$ to $64$, this amounts to $D\approx 80$ to $256$, implying an equal number of Hamiltonian contributions $\hat{H}_{\rm left}[b_l]$  ($\hat{H}_{\rm right}[b_r]$). These describe the action of the Hamiltonian to the left (right) of the sites $i$, $i+1$, while for every $b_l$, $b_c$ ($b_c$, $b_r$) $\hat{h}_{i}[b_l,b_c]$ ($\hat{h}_{i+1}[b_c,b_r]$) is a purely local operator on site $i$ ($i+1$). Combined with the fact that we utilize $\chi=10000$ to $18000$, this large number of Hamiltonian contributions result in memory requirements on the order of several TB for any single ground state calculation and a commensurate amount of computational effort.

Parallelized DMRG  was developed to make calculations of such magnitude tractable~\cite{Bauera}. We provide a more comprehensive overview of its technical features in Appendix~\ref{app:dmrg_tech}, and highlight here its first parallelization layer, which is the most relevant to this work, shown schematically in Fig.~\ref{fig:pdmrg}c: the largest objects of DMRG are the \textit{boundary terms}. These are the expressions of $\hat{H}_{\rm left}[b_l]$, $\hat{H}_{\rm right}[b_r]$ in the $\chi$ optimal basis states of the system to the left or right of sites $i$, $i+1$ respectively. Denoting these basis states as $|\alpha_l\rangle$, $|\alpha_r\rangle$ respectively, the left boundary is $L[b_l]:=\sum_{\alpha_l,\alpha_l'} \left( \langle \alpha_l' | \hat{H}_{\rm left}[b_l] | \alpha_l \rangle \right) |\alpha_l'\rangle  \langle\alpha_l |$, and the right boundary $R$ is defined analogously using $\hat{H}_{\rm right}[b_r]$ and $|\alpha_r\rangle$. The parallelization layer distributes each of the $D$ elements of both these vectors (each element being a block-sparse matrix due to the use of conserved quantum numbers to enhance performance), to nodes of a parallel supercomputer. For this project, the computer cluster was the 'Piz Daint' system of the Swiss National Supercomputing Center (CSCS). In this manner, we have spread the calculations out to up to many dozens of nodes (for the largest lattices).

%%%%%%%%%%%%%%%%%%%%%%%%%%%%%%%%%%%%%%%%%%%%%%%%%%
%%%%%%%%%%%%%%%%%%%%%%%%%%%%%%%%%%%%%%%%%%%%%%%%%%

\section{Scaling analysis}\label{sec:scale}
As discussed in Sec.~\ref{sec:over}, our indirect strategy for arguing that a ground state of the 2D \mbox{U-V} model has SC order, by eliminating competitor phases, relies on obtaining the correct behaviour of the systems spin gap as $\Ltoinf$.
In the following we discuss the two necessary steps required
in service of this goal: sharply reducing the oscillations of the spin gap as $1/L$ decreases, and extrapolating
ground states energy to zero error in the DMRG-approximation.

%%%%%%%%%%%%%%%%%%%%%%%%%%%%%%%%%%%%%%%%%%%%%%%%%%

\subsection{Removing oscillations of $\dsl{L}$ }\label{subsec:remosc}
\begin{figure}[ht]
\centering
\includegraphics[width=1\columnwidth]{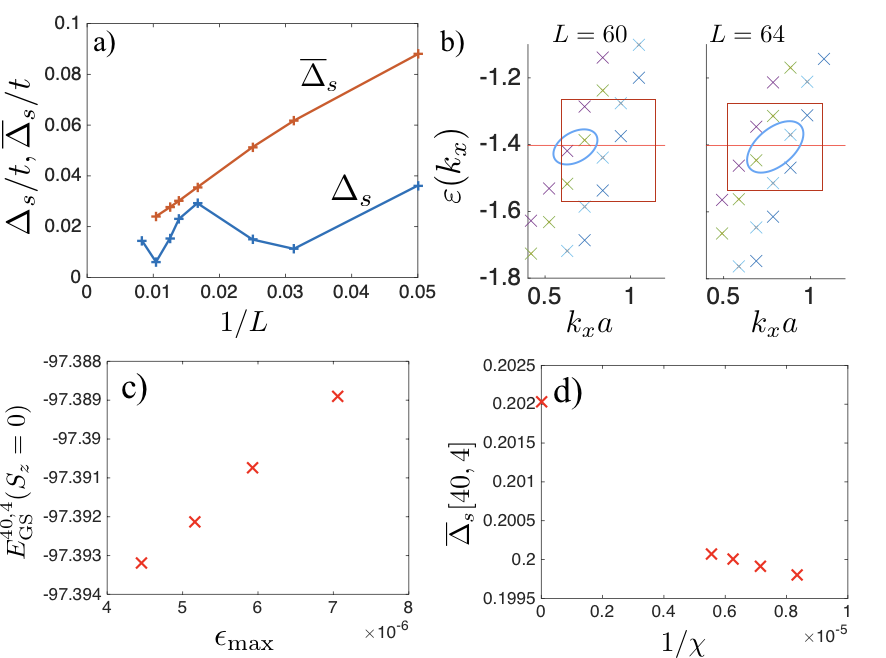}
\caption{
\textbf{(a)} and \textbf{(b)}: Averaging eliminates oscillations in the spin gap.
\textbf{(a)} Comparing $\dsl{L,\Nch=2}$ (blue line) against $\Dsl{L,\Nch=2}$ (red line) for
$U/t=4$, $V/t=1$, $\tp/t=0.1$. Using
the averaged version of the spin gap, $\Ds$, nearly eliminates the oscillations, which are a strong impediment
to a clean extrapolation to $1/L\rightarrow 0$.
\textbf{(b)} Illustrating the cause of oscillatory behaviour in $\ds$ is straightforward for $U=V=0$.
Plotting the single-particle bands for $L=60$ (left), $L=64$ (right) ($\Nch=6$ - only four bands can be seen as two are doubly degenerate each) around the Fermi level (red line), one sees how the distance between highest
occupied/lowest unoccupied levels (pairs marked with blue ellipsis) shifts with $L$, actually increasing here as $L$ increases.
The oscillations vanish almost completely with the averaging over the $\Nch$ bands (red boxes).
\textbf{(c)} and \textbf{(d)}: Scaling ground state energies to zero truncation error $\epsmax$, for the example of $L=40$, $\Nch=4$,
$N_\up=N_\dn=40$.
\textbf{(c)} Plotting $E_{\rm GS}$ against $\epsmax$, computed for $\chi=12000,14000,16000,18000$.
\textbf{(d)} Plotting the resulting $\Dsl{40,4}$ at $1/\chi=0$, and the corresponding values at the four finite
values of $1/\chi$.
}
\label{fig:scalex}
\end{figure}

The standard definition of the spin gap, eq.~(\ref{eq:ds}), is based on the difference between two ground state energies.
However, due to the small value of $\tp/t$ in the regime of interest,  one encounters oscillations of $\dsl{L}$ with $L$,
which presents a complication for obtaining clean extrapolation of $\dsl{L}$ in $L$, as shown in Fig.~\ref{fig:scalex}a.
That weak-interchain coupling is evidently the cause is demonstrated by studying the \mbox{U-V} model in the
non-interacting limit $U=V=0$. While clearly $\ds=0$ for $\Ltoinf$ (at any $\Nch$) in this case,
the $\Nch$ bands are only weakly split. Then, the shifting positions of both the highest occupied as well as of the lowest unoccupied states in each band
with changing $L$ (c.f. Fig.~\ref{fig:scalex}b)
will invariably cause such oscillatory behaviour in the definition~(\ref{eq:ds}).

Based on this analysis, clearly one should smooth this finite-size
effect by using a more general definition of the spin gap. The natural generalization, given by eq.~(\ref{eq:Ds}), is the one
that averages over all the $\Nch$ bands close to the Fermi-level in the non-interacting limit, and that applies unchanged to the interacting regime.
As depicted in Fig.~\ref{fig:scalex}a and b, we find this definition removes the oscillatory behaviour to a large degree
for all our data. As discussed in Appendix~\ref{app:scale}, this definition of the averaged spin gap results in a linear polynomial
in $1/L$ as the only consistent method for extrapolating $\Dsl{L,\Nch}$ to $L^{-1}=0$; linear extrapolation can recover physically expected behaviour correctly, while 
the oscillatory remnant overwhelms any vestiges of quadratic dependency in $1/L$ that might remain after the averaging.

We further note that this procedure is similar to adding two particles instead of one when computing the compressibility of spinless
particles, to avoid the unwanted oscillations provoked by the breaking of $k \to -k$ symmetry that adding a single particle would entail.
Analogously, one usually adds four particles to stay in the sector of total spin zero for spinful systems. 
%BEGIN FIGURE%
%END FIGURE%

%%%%%%%%%%%%%%%%%%%%%%%%%%%%%%%%%%%%%%%%%%%%%%%%%%

\subsection{Extrapolating in the DMRG truncation error}\label{subsec:truncerr}
In DMRG, one attempts to compute the ground state wavefunction $|\psi_{\rm GS}\rangle$
of a lattice-Hamiltonian, by mapping the physical lattice to a 1D chain. At
every bipartition of this DMRG chain, the algorithm achieves the required reduction of the exponentially-scaling
full Hilbert space by approximating $|\psi_{\rm GS}\rangle$ through two
sets of $\chi$ optimally chosen basis states, yielding a wavefunction $|\psi^\chi_{\rm GS}\rangle$. 
A given $\chi$ corresponds to a particular truncation
error $\epsilon:=\langle\delta\psi|\delta\psi\rangle$, ${|\delta\psi\rangle:=|\psi_{\rm GS}\rangle-|\psi^\chi_{\rm GS}\rangle}$,
where $\epsilon$ is known in principle in DMRG.

The controlled accuracy and known error $\epsilon$ enables local quantities, and thus energies,
to be extrapolated to zero error~\cite{White2007}. Specifically for energies, it is known that the approximation
error in the energy, $\delta E$, is $\propto\epsilon$. As we obtain $\Dsl{L,\Nch}$ from the difference
between the $S_z=\Nch$ and $S_z=0$ ground state energies, wherever we have at least two ground states with different $\chi$ we use this technique to extrapolate
the energy to zero $\epsilon$ before computing $\Ds$. In this, we always employ $\epsmax$,
the maximal truncation error committed in the final sweep as a proxy for $\epsilon$. For $V/t=1$, $\Nch=4$ we have
obtained multiple energies, by computing ground states for $\chi=10000$ and $18000$ independently, then
obtained $\chi=12000,14000$ and $16000$ using the $\chi=18000$ ground state as an initial state
and applying two complete sweeps at the lower $\chi$. An example is of this is shown in Fig.~\ref{fig:scalex}c.
As illustrated in Fig.~\ref{fig:scalex}d, the change in $\Dsl{\Ltoinf,\Nch}$ relative to values at finite-$\chi$
is typically small anyway, on the orders of a few percent at most.

Outside these parameters, we have aimed to produce both $\chi=10000$ and $\chi=18000$
ground states in independent simulations wherever possible, but particularly for $V/t=0.5$ available
computing ressources proved ultimately insufficient. Thus, many results for this parameter are based
partly or fully on a single $\chi=10000$ ground state.

In order to obtain any particular value of $\Dsl{L,\Nch}$ as the basis for an extrapolation
to $\Ltoinf$ at fixed $\Nch$, we therefore pursue two separate protocols:

\indent \textbf{Straight extrapolation} - Here, we extrapolate both $S_z=0$ and $S_z=\Nch$
ground state energies to zero $\epsmax$ when possible. If scaling is only possible
in one of the spin sectors, we form $\Ds$ from the two lowest-energy ground
states with comparable $\epsilon$.

\indent \textbf{Extrapolation plus estimates} - Whenever a ground state is available
at $\chi=10000$ exclusively, we estimate the correction heuristically. We have done
this based on the following observations, made on states where extrapolations
were possible: (i) the fractions by which energies further decrease
upon extrapolation seem to roughly double going from one $L$ to the next, (ii) they
also seem to, roughly double with every increase of $\Nch$. (iii) for the same parameters,
fractions seem to be about $1.3$ larger in the $S_z=\Nch$ sector compared to the $S_z=0$ sector.
We then assume an uncertainty of $25\%$ in these estimated correction fractions.

%%%%%%%%%%%%%%%%%%%%%%%%%%%%%%%%%%%%%%%%%%%%%%%%%%
%%%%%%%%%%%%%%%%%%%%%%%%%%%%%%%%%%%%%%%%%%%%%%%%%%

\section{Extrapolated spin gaps and inferring SC-order from correlation functions}\label{sec:res}

The exact behaviour of $\Dsl{L,\Nch}$ as $1/L$ scales to zero is critical to our approach. In~\ref{subsec:spingap},
we take the data as they appear to be, linearly increasing in $1/L$, with only weak remnants of the oscillatory behaviour
that $\Ds$ was defined to eliminate (c.f. Appendix~\ref{app:scale} for details). It then follows that the spin gap is finite for the infinitely long strips, and non-decaying
or even increasing in $\Nch$. In~\ref{subsec:spingap_alt}, we consider the possibility that what appear to be remnants of oscillations
is actually the onset of a scaling of $\Dsl{L,\Nch}$ to zero as $1/L$ decreases. In~\ref{subsec:corrs}, we then analyse the corrollary
information that correlation functions offer.

%%%%%%%%%%%%%%%%%%%%%%%%%%%%%%%%%%%%%%%%%%%%%%%%%%

\subsection{The spin gap as a function of $\Nch$} \label{subsec:spingap}
\begin{table}[t]

    \begin{tabular}{ c || c | c || c | c |}
                            &  \multicolumn{4}{|c|}{ $\Dsl{\Nch}$ [$t\times 10^{-3}$] }     \\ \cline{2-5}
                          &  \multicolumn{2}{|c||}{ Straight extrapolation }    & \multicolumn{2}{|c|}{ Extr. + estimates } \\ \hline
    $\Nch     $ & $V/t=0.5$ & $V/t=1$ & $V/t=0.5$ & $V/t=1$ \\ \hline\hline
    $2$ & $5.59$ & $7.59$  & $5.36$ & $7.31$  \\ \hline
    $4$ & $9.68$  & $10.2$  & $^{(a)}10.77(10)$ & $10.268(11)$ \\ \hline
    $6$ & $9.37$ & $11.4$ & $^{(b)}4.3(1.4)$ &  $11.20(97)$ \\ \hline
    $8$ & $9.45$ & & $^{(c)}7.6(1.3)$ &   \\ \hline
    \end{tabular}
    \caption{
\textbf{(a)} Summary of spin gaps at $\Ltoinf$ for different $\Nch$, assuming the $1/L$-scaling of $\Dsl{L,N}$ we appear to observe continues to hold beyond the strip lengths we computed. ``Extr. + estimates'' is based on estimating
ground state energies at zero truncation error ($\epsilon=0$) for cases where ground states cannot be systematically extrapolated to $\epsilon=0$, because converged simulations were only available for a single $\chi$-value (see text for details). $^{(a)}$:~only one ground state energy of the finite-$L$ ensemble used to obtain this value via linear fit was estimated. $^{(b)}$:~multiple finite-$L$ ground state energies were estimated. $^{(c)}$:~All finite-$L$ ground state energies were estimated.
}
 \label{tab:spingaps}
\end{table}
%END TABLE%
%BEGIN FIGURE%
\begin{figure*}
\includegraphics[width=\textwidth]{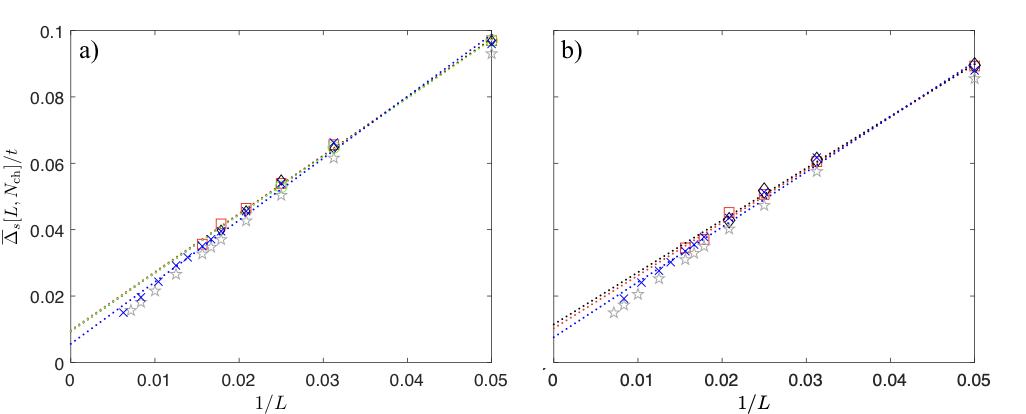}
\caption{
Scaling $\Dsl{L,\Nch}$ to $\Ltoinf$ using the straight extrapolation
protocol (c.f. Sec.~\ref{subsec:truncerr}).
\textbf{(a)} Scaling at $V/t=0.5$, for $\Nch=2$ (blue, crosses), $\Nch=4$ (red, squares), $\Nch=6$ (black, diamonds),
$\Nch=8$ (green, circles).
\textbf{(b)} Scaling at $V/t=1$, symbols and colours for $\Nch=2,4,$ and $6$ as in (a).
In both subfigures, $\dsl{L,\Nch=1}$ has been inserted for comparison purposes (grey stars).
The raw data is summarized in Tab.~\ref{tab:Es_raw} of Appendix~\ref{app:scale}.
}
\label{fig:scalgap}
\end{figure*}
%END FIGURE%
%BEGINFIGURE%
\begin{figure}[h]
\centering
\includegraphics[width=1\columnwidth]{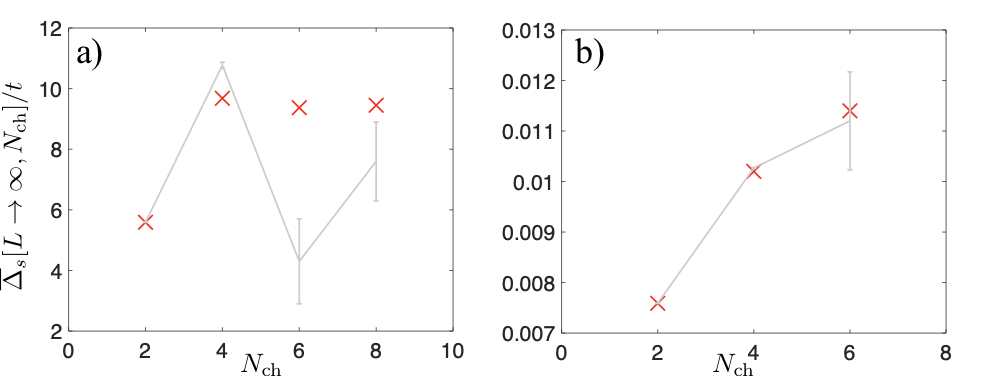}
\caption{
Plotting  $\Dsl{\Ltoinf,\Nch}$ vs. $\Nch$ for
\textbf{(a)}~$V/t=0.5$.
Red crosses denote outcomes of the straight extrapolation protocol, while grey lines with uncertainty ranges
show outcomes of the extrapolation plus estimates protocol (c.f. Sec.~\ref{subsec:truncerr} and Tab.~\ref{tab:spingaps}).
\textbf{(b)}~$V/t=1$. Symbols as in (a).
}
\label{fig:gapNch}
\end{figure}
%END FIGURE%

Applying the two distinct protocols for computing $\Dsl{L,\Nch}$ outlined in Sec.~\ref{subsec:truncerr},
the behaviour of the spin gap under increasing $\Nch$ emerges.
In Figs.~\ref{fig:scalgap}a and b, we show how we obtain spin gaps in the thermodynamic limit $\Ltoinf$
through linear fits, for both $V/t=0.5$ and $V/t=1$, using the straight extrapolation protocol.
Not shown are the equivalent plots for the extrapolation plus estimates protocol. There, we also
employ linear fits, but now we obtain the maximal range of possible outcomes of $\Dsl{\Ltoinf,\Nch}$
by performing separate linear fits for every possible combination of extremal values of the
$\Dsl{L,\Nch}$ at different $L$. The resulting $\Dsl{\Ltoinf,\Nch}$, respectively the mean values
and maximal/minimal values, of both protocols are summarized in Tab.~\ref{tab:spingaps}
and shown in Figs.~\ref{fig:gapNch}a and b.

Using the straight extrapolation protocol, there seems to be little room for doubt that $\Dsl{\Nch}$
is at least a non-decaying function beyond $\Nch=2$ for $V/t=0.5$, while it is unambiguously
a monotonically increasing function at $V/t=1$. At this latter value of the nearest-neighbour repulsion,
the extrapolation plus estimates protocol is benefitting from relatively small uncertainty, as here
we have a greater collection of states that we could extrapolate to $\epsmax=0$, and thus heuristic extrapolations
were only necessary for a few states. With these smaller uncertainties, this second protocol supports the
straight extrapolation protocol. While the non-decaying nature of $\Dsl{\Nch}$ beyond $\Nch=6$ cannot be guaranteed, a scenario
in which the the spin gap suddenly reverses at larger $\Nch$ and starts decaying towards zero
seems highly unlikely and we are not aware of any mechanism that would support such a reversal.
We find further evidence for this view with the significant enhancement of short-range $d_{xy}$-correlations as $\Nch$
increases, which we discuss further in Sec.~\ref{subsec:corrs}.

For $V/t=0.5$ the situation is possibly more uncertain when comparing the $\Dsl{\Ltoinf,\Nch}$
resulting from the two different extrapolation protocols. The deviations are noteworthy for $\Nch=6$, but it is very easily
possible for our heuristic model of the corrections (described in Sec.~\ref{subsec:truncerr}) to be just off in this case.
But we will observe that neither protocol is compatible with a steady decrease of the spin gap with
increasing $\Nch$. And, as for $V/t=1$, a significant and consistent enhancement of short-range
$d_{xy}$ correlations at $\Nch>2$ (see below) further supports this reading of the spin gap behaviour
We thus can summarize that for $V/t=1$ the extended \mbox{U-V} model is highly likely
to have a finite spin gap for $\Nch\rightarrow\infty$, while for $V/t=0.5$ this likelihood,
while reduced, remains high.

%%%%%%%%%%%%%%%%%%%%%%%%%%%%%%%%%%%%%%%%%%%%%%%%%%

\subsection{Testing the alternative: could $\Dsl{\Ltoinf,\Nch}$ scale to zero with $\Nch$?}
\label{subsec:spingap_alt}
%BEGIN FIGURE%
\begin{figure}
\includegraphics[width=1\columnwidth]{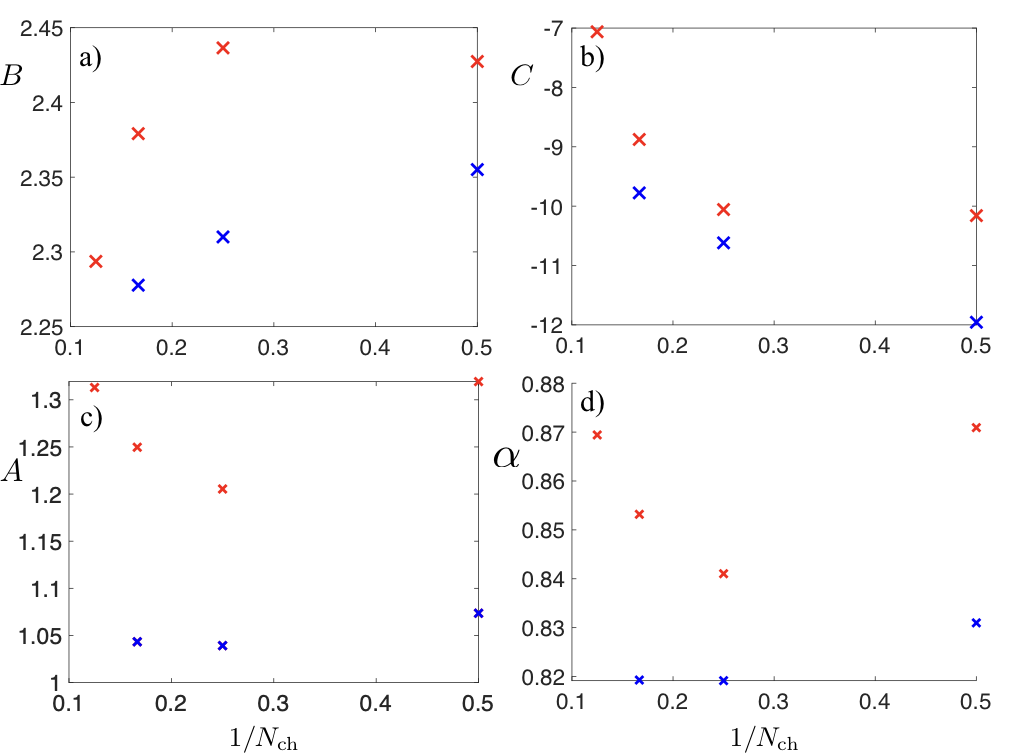}
\caption{
\textbf{(a)} and \textbf{(b)} Testing the viability of a MO/FL state in the 2D limit as an alternative explanation
of the spin gap data, by evaluating the stability of the scaling ansatz ${B[\Nch]/L+C[\Nch]/L^2}$ with $1/\Nch$
for both $V/t=1$ (blue crosses) and $V/t=0.5$ (red crosses).
The strong growth of $C[\Nch]$ indicates the instability of the quadratic term, making the MO/FL-hypothesis
an unsuitable one, except potentially at $V/t=0.5$ (c.f. text for details)
\textbf{(c)} and \textbf{(d)} Testing the viability of a nodal USC state in the 2D limit as an alternative explanation
of the spin gap data, by evaluating the stability of the scaling ansatz ${A[\Nch]/L^{\alpha[\Nch]}}$ with $1/\Nch$
(symbols matching (a) and (b)). For $V/t=1$ there is a quick saturation to seemingly stationary values, indicating that
nodal USC cannot definitely be ruled out as an alternative in the 2D limit (c.f. text for details).
}%
\label{fig:altscalings}
\end{figure}
%END FIGURE%

As discussed, the results for $\Dsl{L,\Nch}$ summarized in Figs.~\ref{fig:scalgap}a,b appear to be most congruent with a linear scaling
in $1/L$ to a finite intersect at $1/L=0$, with weak oscillations surviving the averaging inherent to definition~(\ref{eq:Ds}) (c.f. also Appendix~\ref{app:scale}). 

Returning to the discussion of Sec~\ref{subsec:obs}, the main follow-up question then must be: could there be alternate
fits \textit{that make physical sense}, associated with either nodal superconductivity, or with a Fermi liquid, or a magnetically ordered state?
The first possibility would be marked by $\Dsl{L,\Nch}$ scaling to zero as $A/L^\alpha$, $\alpha<1$ in the \mbox{large-$\Nch$} regime,
while for the latter two possibilities $\Dsl{L,\Nch}$ should also scale to zero in the \mbox{large-$\Nch$} regime, but as $B/L+C/L^2$. We allow for a quadratic correction term
to the ideal linear vanishing of $\Dsl{L,\Nch}$ expected for a FL or MO state (c.f. Sec~\ref{subsec:obs}) because our results from Sec.~\ref{subsec:spingap} show that a
purely linear scaling to zero is impossible given our data.

Analyzing our data under these alternative scenarios requires recognizing that at $\Nch=2$ we start in a regime 
where we are guaranteed a finite spin gap~\cite{BookGiamarchi2003} (c.f. also Appendix~\ref{app:scale}). That makes it extremely likely
that some finite spin gap could survive for at least some larger $\Nch$ - our linear-polynomial ansatz of the previous sections yields just that.

For this reason one cannot just blindly fit the data in Tab.~\ref{tab:Es_raw} with these two alternate scaling ansatzes; these ansatzes will always ``fit" \textit{any} data like in those tables, appearing linear within the fitting region via parameter-tuning, then bending down to zero outside of it. They easily ``fit" even the $\Nch=2$ data, a nonsensical result considering these systems are known to have a finite spin gap.

Thus, the only meaningful question is whether or not these fits converge on a physically consistent scenario as $1/\Nch$ decreases.
For this purpose, we plot the fitting parameters of both ansatzes to the raw data of Tab.~\ref{tab:Es_raw} 
against $1/\Nch$ in Fig.~\ref{fig:altscalings}. For the FL/MO-state ansatz $B[\Nch]/L+C[\Nch]/L^2$ in Figs.~\ref{fig:altscalings}a and b, neither coefficient shows
discernable convergence: $B[\Nch]$ drops monotonically, albeit slowly, while $C[\Nch]$ increases monotonically, at a much faster rate. 
Thus, within the range of accessible $\Nch$-values, the quadratic correction term appears unstable. Rather, the rapid growth of $C[\Nch]$ with growing $\Nch$, coupled with the slow flattening of the linear slope controlled by $B[\Nch]$, is that of an ansatz that increasingly struggles to approximate a behaviour of $\Dsl{L,\Nch}$ that is actually linear or near-linear in $1/L$.

The linear-polynomial ansatz was already treated in Sec.~\ref{subsec:spingap}, yielding a finite spin gap. 
There is a motivated alternative however. A near-linear ansatz $A[\Nch]/L^{\alpha[\Nch]}$ with $\alpha<1$, that scales to zero gap would be suggested by earlier
theory in contradiction to our result. These prior works were based on analytical perturbative RG and mean-field theory; for our 2D U-V model here,
a $d_{x^2-y^2}$-order has been predicted~\cite{Nickel2005,Nickel2006,Bourbonnais2011,Sedeki2012}. Any order other
than isotropic \mbox{$s$-wave} results in nodal lines intersecting with the Fermi surface (c.f.~\ref{fig:ordpar}).
The intersects result in gap-nodes, isolated points in the Brillouin zone where the SC gap closes. 
At these nodes zero-energy quasiparticles may thus be produced, according to the mean-field treatment of these systems, and
these theories thus predict an ungapped spin excitation spectrum. This nodal structure results
in the spin susceptibility, eq.~(\ref{eq:spinsuc}), scaling as $1/L^\alpha$, with $\alpha<1$~\cite{Sigrist1991,Hasegawa1996,Moriya2003,Shinagawa2007,Chubukov2008,Vladimirov2011}.

The results of fitting $A[\Nch]/L^{\alpha[\Nch]}$ to our spin-gap data, depicted in Figs.~\ref{fig:altscalings}c and d, then show a difference between $V/t=0.5$ and $V/t=1$ as $\Nch$ increases. In the latter case both coefficients saturate quickly to nearly stable values. For $\alpha[\Nch]$, this is crucially well below $1$. A ground state with nodal USC in the 2D regime could not be completely ruled out on this basis. To what extent this is an artefact of our data being confined to smaller $L$-values at larger $\Nch$ must remain as a key question to be answered in future work.

For $V/t=0.5$ on the other hand, at $\Nch\geq 4$ we cannot find saturation of the growing $\alpha[\Nch]$ within the accessible $\Nch$. Taken together with the linear polynomial fitting done in Sec.~\ref{subsec:spingap} and the behaviour of $C[\Nch]$ in the FL/MO-scaling ansatz above (Fig.~\ref{fig:altscalings}b), it may not be possible to rule out that the $V/t=0.5$ data could move towards a fully linear scaling to zero in the large $\Nch$-limit. Three conjoined caveats apply however: firstly, $\Nch$, is limited to $\leq 8$ here, it is clearly possible that $\alpha[\Nch]$ stabilizes below $1$ at larger $\Nch$. Secondly, the behaviour of $\alpha[\Nch]$ with $\Nch$ could easily be an artefact of the oscillatory remnant in $1/L$ that $\Dsl{L,\Nch}$ retains despite the averaging, analogous to the discussion in Appendix~\ref{app:scale} for the case of a full second-order polynomial fit to the raw data. Thirdly, the results of the straight extrapolation protocol for the spin gap at $\Ltoinf$ (Tab.~\ref{tab:spingaps}, Fig.~\ref{fig:gapNch}a) would point in the opposite direction.

We thus conclude that our data most strongly supports a fully gapped spin spectrum. As laid out in Appendix~\ref{app:scale}, the averaged spin gap $\Dsl{L,\Nch}$ appears to result in a linear structure in $1/L$ with a superimposed weak oscillatory remnant, of unknown analytical form, with its amplitude slowly decreasing with $1/L$. It is those weak downwards oscillations that the two alternate scaling ansatzes we now studied are, erroneously, susceptible to. A conservative reading of our results would still mandate that at $V/t=1$ we cannot completely rule out a spin excitation spectrum with nodal zeros, which would be compatible with the nodal USC systems described by mean-field theories. At the same time, within the confines of the $\Nch$-values that we have treated in this work, there is no consistent scenario for a FL or MO state at $V/t=1$ due to the apparent instability of the quadratic correction as shown in Fig.~\ref{fig:altscalings}b. 

For $V/t=0.5$ the situation is less clear. The upward growth of $\alpha[\Nch]$ parametrizing the fit to a nodal USC-regime, combined with the instability of the quadratic correction term $C[\Nch]$ in the fit to the FL/MO-regime opens up the possibility that there may not be enough data to reliably exclude any of the three regimes for this parameter, when viewed in conjunction with the outcome of the extrapolation plus estimates protocol for the linear-polynomial extrapolation (Fig.~\ref{fig:gapNch}a and Tab.~\ref{tab:spingaps}).
However, viewed together with the enhancement exclusive to the $d_{xy}$-correlation channel discussed in the next section, it turns out that either the nodal USC scenario or the fully spin-gapped one should be favored for $V/t=0.5$ as well.

%%%%%%%%%%%%%%%%%%%%%%%%%%%%%%%%%%%%%%%%%%%%%%%%%%

\subsection{Correlation functions: pairing in the $d_{xy}$-channel as likely candidate-order in the 2D limit}\label{subsec:corrs}
%BEGIN TABLE%
\begin{table*}[t]
   \begin{tabular}{ c || c | c || c | c || c | c || c ||}
                          &  \multicolumn{2}{|c||}{ $\Nch=2$ }    & \multicolumn{2}{|c||}{ $\Nch=4$ } &  \multicolumn{2}{|c||}{ $\Nch=6$ }    & \multicolumn{1}{|c||}{ $\Nch=8$ }\\ \hline
    \mbox{short-range decay }      & $V/t=0.5$ & $V/t=1$ & $V/t=0.5$ & $V/t=1$ & $V/t=0.5$ & $V/t=1$ & $V/t=0.5$  \\ \hline\hline
    $\mbox{\Large\( \frac{\max_{|r-r_s|\leq 1}[d_{xy}(r)]}{d_{xy}(1)} \)}$                  & $0.17$ & $0.15$   & \cellcolor{red!25} $1.08$ &  \cellcolor{red!25} $0.55$ &  \cellcolor{red!25} $0.67$ &  \cellcolor{red!25} $0.79$  &  \cellcolor{red!25} $1.05$  \\ \hline
    $\mbox{\Large\( \frac{\max_{|r-r_s|\leq 1}[d_{x^2-y^2}(r)]}{d_{x^2-y^2}(1)} \)}$ & $0.09$ & $0.08$  & $0.08$ & $0.09$ & $0.09$ & $0.07$  & $0.09$  \\ \hline
    $\mbox{\Large\( \frac{\max_{|r-r_s|\leq 1}[s(r)]}{s(1)} \)}$                                  & $0.97$ & $0.87$  & $0.22$ & $0.09$ & $0.10$ & $0.16$  & $0.14$  \\ \hline
    $\mbox{\Large\( \frac{\max_{|r-r_s|\leq 1}[C(r)]}{C(1)} \)}$                                 & $0.06$ & $0.07$  & $0.06$ & $0.06$ & $0.06$ & $0.06$  & $0.06$ \\ \hline

    \end{tabular}
\caption{Decay of correlations at short range, $r\approx r_s=4a$, in the effective 2D regime (c.f. main text and example shown in Fig,~\ref{fig:corrdecay}a). The strong enhancement in the $d_{xy}$-channel compared to all other channels for $\Nch\geq 4$ is highlighted in red. }%
  \label{tab:corrdecay}
\end{table*}
%END TABLE%

%BEGIN FIGURE%
\begin{figure}
\includegraphics[width=1\columnwidth]{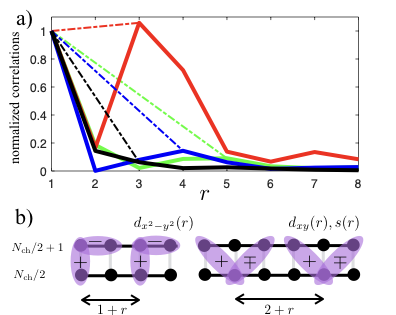}
\caption{
\textbf{(a)} Correlation functions for $\Nch=8$, $V/t=0.5$, normalized to their value at $r=1$. Dash-dotted lines show relative decay at first maximum within $r\lesssim r_s$ for $d_{xy}$, (red), $d_{x^2-y^2}$ (green), $s$ (blue) and $C$ (black). This behaviour of the normalized correlation functions in representative for all $\Nch\geq 4$.
\textbf{(b)}~How correlation functions $d_{x^2-y^2}$, $d_{xy}$ and $s(r)$, defined in eq.~(\ref{eq:dx2y2corr})~-~(\ref{eq:ddcorr}), are evaluated on the the two
central legs of the lattice. 
}%
\label{fig:corrdecay}
\end{figure}
%END FIGURE%

From the ground states, any desired correlation function between site ${\bf r}_1$ and site ${\bf r}_2$ in the lattice can in principle be computed. With the form-factor of the lattices we consider, strips of length $L$ and width $\Nch$, where $L\gg\Nch$, generally a cross-over of behaviours is to be expected with growing $r:=|{\bf r}_1-{\bf r}_2|$. As $\Nch$ increases, at short distances the physics will become increasingly dominated by that of the 2D limit ($\Nch\rightarrow\infty$). Conversely, at sufficiently large distances, any correlation function will exhibit the typical behaviour of a 1D-system, i.e. it will decay algebraically as a function of $r$ - as long as no discrete symmetry breaks spontaneously~\footnote{a transition to a CDW would happen in our systems eventually if $V/t$ were to be increased further}. This behaviour is generic to the ground state of any 1D quantum system, even one of finite width~\cite{BookGiamarchi2003}. The length-scale at which the cross-over between the two regimes takes place will increase as $\Nch$ is raised, with the 1D-regime starting at larger and larger distances, until it disappears completely when $\Nch\rightarrow\infty$.

Even with pDMRG we cannot access the full 2D regime for the time being, but we can get new and valuable insight into how the various potential channels for ordering of the \mbox{U-V} model in the 2D limit respond to the systematic increase of $\Nch$. As we already evaluate the possibility of magnetic order or a Fermi liquid based on the scaling of $\Dsl{L,\Nch}$, we limit ourselves to the main competitor-orders, the SC $d_{x^2-y^2}(r)$-, $d_{xy}(r)$-, $s(r)$- as well as the charge-density-wave channel. We compute the correlation functions~(\ref{eq:dx2y2corr})~-~(\ref{eq:ddcorr}) connected to each of these, as defined in Sec.~\ref{subsec:obs}.

We then specifically study the dependency of the short-range behaviour, $r \lesssim r_s:=\pi / \kf=4a$ of the above correlation functions on $\Nch$. An example that is highly representative of the general behaviour for $\Nch\geq 4$ is shown in Fig.~\ref{fig:corrdecay}: when each correlation function is normalized to its value at $r=1$, the first maximum within the short-distance quasi-2D regime decays strongly in all channels \textit{except} in the $d_{xy}$ channel. Table~\ref{tab:corrdecay}, which summarizes the short-range decay for all studied $\Nch$-values, confirms this short-range enhancement as exclusive to the $d_{xy}$ channel. As is to be expected, once $r$ increases beyond the short-range regime, $d_{xy}(r)$, like all the other correlation functions, behaves according to 1D physics, i.e. it decays algebraically. 

We do not take the above short-range behaviour of these systems at $\Nch\leq 8$ to constitute definite proof of $d_{xy}$-ordering in the full 2D-limit. At the same time however, we observe that it would align with a simple order-of-magnitude estimate, namely that the diagonal singlet pairing of $d_{xy}$-order is  energetically advantageous at strong coupling - naive perturbative analysis suggests that the gain in energy should be $\mathcal{O}(\tp^2/V)$ across a diagonal, while it is $\mathcal{O}(\tp^2/U)$ for singlet pairing across a rung in $d_{x^2-y^2}$ order. And of course, this would be in line with the high likelihood for a finite spin gap and the attendant elimination of competing MO or FL phases in these systems that we discussed in Secs.~\ref{subsec:spingap} and \ref{subsec:spingap_alt}; the open questions raised by a system having a finite spin gap jointly with SC $d_{xy}$-order we will discuss now.

%%%%%%%%%%%%%%%%%%%%%%%%%%%%%%%%%%%%%%%%%%%%%%%%%%
%%%%%%%%%%%%%%%%%%%%%%%%%%%%%%%%%%%%%%%%%%%%%%%%%%

\section{Discussion and connections to experiment}\label{sec:disc}

Our results show a fully gapped spin excitation spectrum for finite-width strips of the 2D \mbox{U-V} model at the given
parameters, as evidenced by the finite ${\Dsl{\Ltoinf,\Nch}}$ we find. 

We then show that the behaviour of the gap with the strip-width is most consistent with a fully-gapped spin
spectrum also in the 2D limit when taking the apparent structure of $\Dsl{L,\Nch}$ as consisting of a linear part
with a superimposed weak oscillatory remnant into account (c.f. Sec.~\ref{subsec:spingap_alt} and Appendix~\ref{app:scale}). 
At the same time, the behaviour of the various correlation functions at short range
with increasing width, indicative of the behaviour of the 2D system, lends support to pairing happening predominantly 
in the $d_{xy}$-channel, rather than in the extended $s$- or $d_{x^2-y^2}$-channel.

\begin{figure}[h]
\centering
\includegraphics[width=0.9\columnwidth]{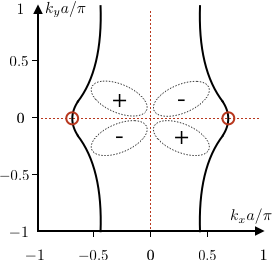}
\caption{
Fermi surface of the noninteracting lattice electrons for the 2D \mbox{U-V} model (solid black lines) and 
possible symmetry of the order parameter, shown as a generic mixture between $d_{xy}$ and $g_{xy(x^2-y^2)}$
(see text for details). This symmetry minimizes, but does not eliminate intersections of nodal lines and Fermi surface
(red circles), where the gap of USC order would exhibit zero-nodes in standard mean field treatments. From this, these theories
predict a non-gapped spectrum for spin excitation, i.e. zero spin gap.
}
\label{fig:ordpar}
\end{figure}

Such a conclusion however is at variance with existing analytical descriptions of USC systems, such as from combining
the results of perturbative functional RG with mean-field descriptions. Besides these methods predicting $d_{x^2-y^2}$
symmetry for the order parameter instead~\cite{Nickel2005,Nickel2006,Bourbonnais2011,Sedeki2012}, the more far-reaching difference to our findings lies in the effects these theories predict from the nodal lines,
which any plausible order parameter will exhibit (these originate from the systems attempt to minimize the Coulomb repulsion).
Specifically, they forecast that node lines' intersections with the Fermi surface will always create point nodes for 
gapless quasiparticle excitations (c.f. Fig.~\ref{fig:ordpar}). In the mean-field theory, these point nodes in turn necessitate an ungapped spin excitation spectrum.

There are two ways to look at the apparent discrepancy between this mean-field treatment and our findings of a finite spin gap in Sec.~\ref{subsec:spingap}. 
We considered the first way in Sec.~\ref{subsec:spingap_alt}, observing that disregarding the apparent structure of $\Dsl{L,\Nch}$ (linear with oscillatory remnant, c.f. Appendix~\ref{app:scale}) allows
for sublinear scaling of $\Dsl{L,\Nch}$ to be a physically consistent fit within the accessible $\Nch$-range, at least for $V/t=1$. As explained there, this scaling of 
$\Dsl{L,\Nch}$ taken together with the boost to the $d_{xy}$-correlations from Sec.~\ref{subsec:corrs} would be consistent with the prediction of nodal USC derived from mean-field theory just the exact symmetry of the order parameter would be different.

The second way is to consider the structure of the $\Dsl{L,\Nch}$-data as linear + weak oscillations. Then one has to reconcile the finite spin gap in the 2D limit that ultimately results 
from our numerics at strong coupling with the weak-coupling analytical theory. Taking stock of the experimental situation allows just such a reconciliation. Firstly, we notice that our result of dominant $d_{xy}$ correlations in the 2D regime of the strips' ground state 
would be in line with the general argument that it is energetically advantageous for the system to minimize the number of nodes, which $d_{xy}$ would achieve better than a
$d_{x^2-y^2}$-order given the Fermi-surface topology (c.f. Fig.~\ref{fig:ordpar}). Secondly, measurements on actual USC materials, which are certainly in the 
strongly interacting regime, provide evidence for USC order with zero-gap point nodes, yet which exhibit \textit{fully gapped} spin excitations at the same time. This has been demonstrated for LSCO~\cite{Lake1999} close to and away from optimal doping, as well as YBCO~\cite{Dai2001} over a range of doping. To our knowledge, this fundamental discrepancy between the standard mean-field descriptions of USC models with zero-gap nodes and actual measurements so far has neither been theoretically explained nor even replicated. Thirdly, strongly underdoped yet still superconducting LSCO
has been shown to have not just gapped spin excitations but to be fully gapped overall~\cite{Razzoli2013}. The possibilities discussed for this observation are either another phase
coexisting with USC order or topological superconductivity of the chiral $d_{x^2-y^2}\pm id_{xy}$-type, which the authors show to fit the measured gap. Fourthly, the loss of
$D_4$ symmetry in the anisotropic 2D lattice results in a mixing of orbitals, such as $d_{xy}$ and $g_{xy(x^2-y^2)}$ (c.f. Fig.~\ref{fig:ordpar})~\cite{Cho2013}. When viewed in light of
the numerous proposals to explain USC phases at low or zero temperature via mixing of different order symmetries, it cannot be ruled out that the remaining nodes resulting from a pure 
$d_{xy}+\alpha g_{xy(x^2-y^2)}$ orbital, sketched in Fig.~\ref{fig:ordpar}, are eliminated when admixed with yet other orders, even without invoking the possibility of topological superconductivity.

Given the available data we cannot currently distinguish among these possibilities, but provide them to make it plain
that the most consistent interpretation of our results, a fully spin-gapped USC order (with dominant $d_{xy}$ symmetry), has clear precedents,
the predictions of weak-coupling analytics notwithstanding.
%BEGIN FIGURE%
%END FIGURE%

%%%%%%%%%%%%%%%%%%%%%%%%%%%%%%%%%%%%%%%%%%%%%%%%%%
%%%%%%%%%%%%%%%%%%%%%%%%%%%%%%%%%%%%%%%%%%%%%%%%%%

\section{Conclusion}\label{sec:out}

We have performed large-scale calculations with parallel DMRG to obtain ground states
of the 2D \mbox{U-V} Hubbard model, the presumed minimal model of unconventional superconductivity
in the organic Bechgaard and Fabre salts. The use of parallel DMRG here allows to study large systems at sufficient
accuracies to reliably compute ground state energy differences between different quantum number sectors, i.e. the spin gap.

Considering strips of increasing width up to $8$ chains, we have computed
this models' spin gap as a function of strip-length and -width, averaged so as to strongly reduce oscillations that
would preclude extrapolations to infinite strip-length. After extrapolating the ground state energies in the discarded weight, 
we arrived at data consistent with a finite spin gap at the thermodynamic limit (i.e. at infinite length). Taking the apparent structure
of the computed spin gaps as a function of inverse length into account, we conclude that the spin gap of the infinite strips 
is non-decreasing or even increasing with the strips width. A finite spin gap in the full 2D limit is therefore the most likely
possibility, thus making the case for singlet pairing. Disregarding key features of the computed spin gaps (i.e. its decomposition into
linear and weak oscillatory remnant) still yields scaling results at most consistent with nodal unconventional superconductivity, 
where the spin excitation spectrum would touch zero in isolated points. Even then however our results do not 
admit a physically consistent scaling scaling solution in line with a Fermi liquid or magnetically ordered states.

As we steer away from model parameters that could result in charge density wave order of the ground state, the remaining
candidate by exclusion is unconventional (i.e. repulsively mediated) superconducting order of singlet type. This reading is further supported by the
enhancement of short range correlations (i.e. where the strips' resemble a 2D system) with the strips' width to take place only in the $d_{xy}$-channel.

These results suggest several lines for further inquiry. One would be to test for the possibility of topological $d_{x^2-y^2}\pm id_{xy}$ superconductivity
that is raised by the presence of the spin gap at strong coupling.
Another would be to move more closely to the parameter regime of the Bechgaard and Fabre salts, i.e. to increase both $U/t$ and $V/t$ further yet. This will have to entail
a careful study of the enlarged parameter space for the onset of the Mott-insulating state that we discussed in Sec.~\ref{subsec:parms} for the case $\Nch=2$, and the behaviour of which at $\Nch>2$ is currently unclear.

%%%%%%%%%%%%%%%%%%%%%%%%%%%%%%%%%%%%%%%%%%%%%%%%%%
%%%%%%%%%%%%%%%%%%%%%%%%%%%%%%%%%%%%%%%%%%%%%%%%%%

\section{Acknowledgements}\label{sec:ack}
A.K. would like to thank Ronny Thomale for useful discussions.
This work was supported in part by the Swiss NSF under MaNEP and Division II, and has further received funding through an ERC Starting Grant from the European
Union's Horizon 2020 research and innovation programme under grant agreement No. 758935. The development
of pDMRG was made possible by support of the Swiss government under its HP2C initiative. 
Carrying out the large-scale ground state calculations was made possible through two grants of compute time
on the ``Piz Daint'' cluster of the Swiss National Supercomputing Center
(CSCS), proposal IDs 20131011-18256183-84 and 20141010-173654696-445. The authors
would like to thank CSCS for supporting this work. Calculations for $\Nch=1$ and for some values at $\Nch=2$
were performed on resources provided by the Swedish National Infrastructure for Computing (SNIC) at the PDC, KTH Stockholm.

%%%%%%%%%%%%%%%%%%%%%%%%%%%%%%%%%%%%%%%%%%%%%%%%%%
%%%%%%%%%%%%%%%%%%%%%%%%%%%%%%%%%%%%%%%%%%%%%%%%%%

\bibliography{/Users/kantian/Documents/library.bib}

\begin{thebibliography}{67}
\expandafter\ifx\csname natexlab\endcsname\relax\def\natexlab#1{#1}\fi
\expandafter\ifx\csname bibnamefont\endcsname\relax
  \def\bibnamefont#1{#1}\fi
\expandafter\ifx\csname bibfnamefont\endcsname\relax
  \def\bibfnamefont#1{#1}\fi
\expandafter\ifx\csname citenamefont\endcsname\relax
  \def\citenamefont#1{#1}\fi
\expandafter\ifx\csname url\endcsname\relax
  \def\url#1{\texttt{#1}}\fi
\expandafter\ifx\csname urlprefix\endcsname\relax\def\urlprefix{URL }\fi
\providecommand{\bibinfo}[2]{#2}
\providecommand{\eprint}[2][]{\url{#2}}

\bibitem[{\citenamefont{Ziman}(1972)}]{BookZiman1972}
\bibinfo{author}{\bibfnamefont{J.~M.} \bibnamefont{Ziman}},
  {\it\bibinfo{title}{{Principles of the Theory of Solids}}}
  (\bibinfo{publisher}{Cambridge University Press},
  \bibinfo{address}{Cambridge}, \bibinfo{year}{1972}).

\bibitem[{\citenamefont{Giamarchi}(2003)}]{BookGiamarchi2003}
\bibinfo{author}{\bibfnamefont{T.}~\bibnamefont{Giamarchi}},
  {\it\bibinfo{title}{{Quantum Physics in One Dimension}}}
  (\bibinfo{publisher}{Oxford University Press}, \bibinfo{year}{2003}),
  \bibinfo{edition}{1st} ed..

\bibitem[{\citenamefont{Anderson}(1997)}]{BookAnderson1997}
\bibinfo{author}{\bibfnamefont{P.~W.} \bibnamefont{Anderson}},
  {\it\bibinfo{title}{{The Theory of Superconductivity in the High-T{\_}c
  Cuprate Superconductors}}} (\bibinfo{publisher}{Princeton University Press},
  \bibinfo{address}{Princeton, NJ}, \bibinfo{year}{1997}).

\bibitem[{\citenamefont{Orenstein}(2000)}]{Orenstein2000}
\bibinfo{author}{\bibfnamefont{J.}~\bibnamefont{Orenstein}},
  \bibinfo{journal}{Science} \textbf{\bibinfo{volume}{288}},
  \bibinfo{pages}{468} (\bibinfo{year}{2000}).

\bibitem[{\citenamefont{Varma}(1985)}]{Varma1985}
\bibinfo{author}{\bibfnamefont{C.~M.} \bibnamefont{Varma}},
  \bibinfo{journal}{Comments Solid State Phys.} \textbf{\bibinfo{volume}{11}},
  \bibinfo{pages}{221} (\bibinfo{year}{1985}).

\bibitem[{\citenamefont{Si and Steglich}(2010)}]{Si2010}
\bibinfo{author}{\bibfnamefont{Q.}~\bibnamefont{Si}} \bibnamefont{and}
  \bibinfo{author}{\bibfnamefont{F.}~\bibnamefont{Steglich}},
  \bibinfo{journal}{Science} \textbf{\bibinfo{volume}{329}},
  \bibinfo{pages}{1161} (\bibinfo{year}{2010}).

\bibitem[{\citenamefont{Bourbonnais and J{\'{e}}rome}(2007)}]{Bourbonnais2007}
\bibinfo{author}{\bibfnamefont{C.}~\bibnamefont{Bourbonnais}} \bibnamefont{and}
  \bibinfo{author}{\bibfnamefont{D.}~\bibnamefont{J{\'{e}}rome}}, in
  {\it\bibinfo{booktitle}{The Physics of Organic Superconductors and
  Conductors}}, edited by
  \bibinfo{editor}{\bibfnamefont{A.}~\bibnamefont{Lebed}}
  (\bibinfo{publisher}{Springer}, \bibinfo{year}{2007}), p. \bibinfo{pages}{pp.
  358}.

\bibitem[{\citenamefont{J{\'{e}}rome}(2012)}]{Jerome2012}
\bibinfo{author}{\bibfnamefont{D.}~\bibnamefont{J{\'{e}}rome}},
  \bibinfo{journal}{J. Supercond. Nov. Magn.} \textbf{\bibinfo{volume}{25}},
  \bibinfo{pages}{633} (\bibinfo{year}{2012}).

\bibitem[{\citenamefont{Damascelli and Shen}(2003)}]{Damascelli2003}
\bibinfo{author}{\bibfnamefont{A.}~\bibnamefont{Damascelli}} \bibnamefont{and}
  \bibinfo{author}{\bibfnamefont{Z.-X.} \bibnamefont{Shen}},
  \bibinfo{journal}{Rev. Mod. Phys.} \textbf{\bibinfo{volume}{75}},
  \bibinfo{pages}{473} (\bibinfo{year}{2003}).

\bibitem[{\citenamefont{Fischer et~al.}(2007)\citenamefont{Fischer, Kugler,
  Maggio-Aprile, Berthod, and Renner}}]{Fischer2007}
\bibinfo{author}{\bibfnamefont{{O}.}~\bibnamefont{Fischer}},
  \bibinfo{author}{\bibfnamefont{M.}~\bibnamefont{Kugler}},
  \bibinfo{author}{\bibfnamefont{I.}~\bibnamefont{Maggio-Aprile}},
  \bibinfo{author}{\bibfnamefont{C.}~\bibnamefont{Berthod}}, \bibnamefont{and}
  \bibinfo{author}{\bibfnamefont{C.}~\bibnamefont{Renner}},
  \bibinfo{journal}{Rev. Mod. Phys.} \textbf{\bibinfo{volume}{79}},
  \bibinfo{pages}{353} (\bibinfo{year}{2007}).

\bibitem[{\citenamefont{Scalapino}(2012)}]{Scalapino2012a}
\bibinfo{author}{\bibfnamefont{D.~J.} \bibnamefont{Scalapino}},
  \bibinfo{journal}{Rev. Mod. Phys.} \textbf{\bibinfo{volume}{84}},
  \bibinfo{pages}{1383} (\bibinfo{year}{2012}).

\bibitem[{\citenamefont{{Varma C. M.} and Giamarchi}(1991)}]{Varma1991}
\bibinfo{author}{\bibnamefont{{Varma C. M.}}} \bibnamefont{and}
  \bibinfo{author}{\bibfnamefont{T.}~\bibnamefont{Giamarchi}}
  (\bibinfo{publisher}{Elsevier}, \bibinfo{year}{1991}).

\bibitem[{\citenamefont{Salje et~al.}(1996)\citenamefont{Salje, Alexandrov,
  Liang, and {Cambridge University Press.}}}]{BookSalje1996}
\bibinfo{author}{\bibfnamefont{E.~K.~H.} \bibnamefont{Salje}},
  \bibinfo{author}{\bibfnamefont{A.~S.} \bibnamefont{Alexandrov}},
  \bibinfo{author}{\bibfnamefont{W.~Y.} \bibnamefont{Liang}}, \bibnamefont{and}
  \bibinfo{author}{\bibnamefont{{Cambridge University Press.}}},
  {\it\bibinfo{title}{{Polarons and Bipolarons in High-Tc Superconductors and
  Related Materials}}} (\bibinfo{publisher}{Cambridge University Press},
  \bibinfo{year}{1996}).

\bibitem[{\citenamefont{Schulz}(1987)}]{Schulz1987}
\bibinfo{author}{\bibfnamefont{H.~J.} \bibnamefont{Schulz}},
  \bibinfo{journal}{Europhys. Lett.} \textbf{\bibinfo{volume}{4}},
  \bibinfo{pages}{609} (\bibinfo{year}{1987}).

\bibitem[{\citenamefont{Furukawa et~al.}(1998)\citenamefont{Furukawa, Rice, and
  Salmhofer}}]{Furukawa1998}
\bibinfo{author}{\bibfnamefont{N.}~\bibnamefont{Furukawa}},
  \bibinfo{author}{\bibfnamefont{T.~M.} \bibnamefont{Rice}}, \bibnamefont{and}
  \bibinfo{author}{\bibfnamefont{M.}~\bibnamefont{Salmhofer}},
  \bibinfo{journal}{Phys. Rev. Lett.} \textbf{\bibinfo{volume}{81}},
  \bibinfo{pages}{3195} (\bibinfo{year}{1998}).

\bibitem[{\citenamefont{L{\"{a}}uchli et~al.}(2004)\citenamefont{L{\"{a}}uchli,
  Honerkamp, and Rice}}]{Lauchli2004}
\bibinfo{author}{\bibfnamefont{A.}~\bibnamefont{L{\"{a}}uchli}},
  \bibinfo{author}{\bibfnamefont{C.}~\bibnamefont{Honerkamp}},
  \bibnamefont{and} \bibinfo{author}{\bibfnamefont{T.~M.} \bibnamefont{Rice}},
  \bibinfo{journal}{Phys. Rev. Lett.} \textbf{\bibinfo{volume}{92}},
  \bibinfo{pages}{037006} (\bibinfo{year}{2004}).

\bibitem[{\citenamefont{Deng et~al.}(2015)\citenamefont{Deng, Kozik, Prokof'ev,
  and Svistunov}}]{Deng2015}
\bibinfo{author}{\bibfnamefont{Y.}~\bibnamefont{Deng}},
  \bibinfo{author}{\bibfnamefont{E.}~\bibnamefont{Kozik}},
  \bibinfo{author}{\bibfnamefont{N.~V.} \bibnamefont{Prokof'ev}},
  \bibnamefont{and} \bibinfo{author}{\bibfnamefont{B.~V.}
  \bibnamefont{Svistunov}}, \bibinfo{journal}{Europhys. Lett.}
  \textbf{\bibinfo{volume}{110}}, \bibinfo{pages}{57001}
  (\bibinfo{year}{2015}).

\bibitem[{\citenamefont{Nickel et~al.}(2005)\citenamefont{Nickel, Duprat,
  Bourbonnais, and Dupuis}}]{Nickel2005}
\bibinfo{author}{\bibfnamefont{J.~C.} \bibnamefont{Nickel}},
  \bibinfo{author}{\bibfnamefont{R.}~\bibnamefont{Duprat}},
  \bibinfo{author}{\bibfnamefont{C.}~\bibnamefont{Bourbonnais}},
  \bibnamefont{and} \bibinfo{author}{\bibfnamefont{N.}~\bibnamefont{Dupuis}},
  \bibinfo{journal}{Phys. Rev. Lett.} \textbf{\bibinfo{volume}{95}},
  \bibinfo{pages}{247001} (\bibinfo{year}{2005}).

\bibitem[{\citenamefont{Nickel et~al.}(2006)\citenamefont{Nickel, Duprat,
  Bourbonnais, and Dupuis}}]{Nickel2006}
\bibinfo{author}{\bibfnamefont{J.~C.} \bibnamefont{Nickel}},
  \bibinfo{author}{\bibfnamefont{R.}~\bibnamefont{Duprat}},
  \bibinfo{author}{\bibfnamefont{C.}~\bibnamefont{Bourbonnais}},
  \bibnamefont{and} \bibinfo{author}{\bibfnamefont{N.}~\bibnamefont{Dupuis}},
  \bibinfo{journal}{Phys. Rev. B} \textbf{\bibinfo{volume}{73}},
  \bibinfo{pages}{165126} (\bibinfo{year}{2006}).

\bibitem[{\citenamefont{Bourbonnais and Sedeki}(2011)}]{Bourbonnais2011}
\bibinfo{author}{\bibfnamefont{C.}~\bibnamefont{Bourbonnais}} \bibnamefont{and}
  \bibinfo{author}{\bibfnamefont{A.}~\bibnamefont{Sedeki}},
  \bibinfo{journal}{C. R. Phys.} \textbf{\bibinfo{volume}{12}},
  \bibinfo{pages}{532} (\bibinfo{year}{2011}).

\bibitem[{\citenamefont{Sedeki et~al.}(2012)\citenamefont{Sedeki, Bergeron, and
  Bourbonnais}}]{Sedeki2012}
\bibinfo{author}{\bibfnamefont{A.}~\bibnamefont{Sedeki}},
  \bibinfo{author}{\bibfnamefont{D.}~\bibnamefont{Bergeron}}, \bibnamefont{and}
  \bibinfo{author}{\bibfnamefont{C.}~\bibnamefont{Bourbonnais}},
  \bibinfo{journal}{Phys. Rev. B} \textbf{\bibinfo{volume}{85}},
  \bibinfo{pages}{165129} (\bibinfo{year}{2012}).

\bibitem[{\citenamefont{Metzner et~al.}(2012)\citenamefont{Metzner, Salmhofer,
  Honerkamp, Meden, and Sch{\"{o}}nhammer}}]{Metzner2012a}
\bibinfo{author}{\bibfnamefont{W.}~\bibnamefont{Metzner}},
  \bibinfo{author}{\bibfnamefont{M.}~\bibnamefont{Salmhofer}},
  \bibinfo{author}{\bibfnamefont{C.}~\bibnamefont{Honerkamp}},
  \bibinfo{author}{\bibfnamefont{V.}~\bibnamefont{Meden}}, \bibnamefont{and}
  \bibinfo{author}{\bibfnamefont{K.}~\bibnamefont{Sch{\"{o}}nhammer}},
  \bibinfo{journal}{Rev. Mod. Phys.} \textbf{\bibinfo{volume}{84}},
  \bibinfo{pages}{299} (\bibinfo{year}{2012}).

\bibitem[{\citenamefont{LeBlanc et~al.}(2015)\citenamefont{LeBlanc, Antipov,
  Becca, Bulik, Chan, Chung, Deng, Ferrero, Henderson, Jim{\'{e}}nez-Hoyos
  et~al.}}]{LeBlanc2015}
\bibinfo{author}{\bibfnamefont{J.~P.~F.} \bibnamefont{LeBlanc}},
  \bibinfo{author}{\bibfnamefont{A.~E.} \bibnamefont{Antipov}},
  \bibinfo{author}{\bibfnamefont{F.}~\bibnamefont{Becca}},
  \bibinfo{author}{\bibfnamefont{I.~W.} \bibnamefont{Bulik}},
  \bibinfo{author}{\bibfnamefont{G.~K.-L.} \bibnamefont{Chan}},
  \bibinfo{author}{\bibfnamefont{C.-M.} \bibnamefont{Chung}},
  \bibinfo{author}{\bibfnamefont{Y.}~\bibnamefont{Deng}},
  \bibinfo{author}{\bibfnamefont{M.}~\bibnamefont{Ferrero}},
  \bibinfo{author}{\bibfnamefont{T.~M.} \bibnamefont{Henderson}},
  \bibinfo{author}{\bibfnamefont{C.~A.} \bibnamefont{Jim{\'{e}}nez-Hoyos}},
  \bibnamefont{et~al.}, \bibinfo{journal}{Phys. Rev. X}
  \textbf{\bibinfo{volume}{5}}, \bibinfo{pages}{041041} (\bibinfo{year}{2015}).

\bibitem[{\citenamefont{Schollw{\"{o}}ck}(2011)}]{Schollwock2011}
\bibinfo{author}{\bibfnamefont{U.}~\bibnamefont{Schollw{\"{o}}ck}},
  \bibinfo{journal}{Ann. Phys.} \textbf{\bibinfo{volume}{326}},
  \bibinfo{pages}{96} (\bibinfo{year}{2011}).

\bibitem[{\citenamefont{Stoudenmire and White}(2012)}]{Stoudenmire2012}
\bibinfo{author}{\bibfnamefont{E.}~\bibnamefont{Stoudenmire}} \bibnamefont{and}
  \bibinfo{author}{\bibfnamefont{S.~R.} \bibnamefont{White}},
  \bibinfo{journal}{Annual Review of Condensed Matter Physics}
  \textbf{\bibinfo{volume}{3}}, \bibinfo{pages}{111} (\bibinfo{year}{2012}).

\bibitem[{\citenamefont{Carlson et~al.}(2000)\citenamefont{Carlson, Orgad,
  Kivelson, and Emery}}]{Carlson2000}
\bibinfo{author}{\bibfnamefont{E.~W.} \bibnamefont{Carlson}},
  \bibinfo{author}{\bibfnamefont{D.}~\bibnamefont{Orgad}},
  \bibinfo{author}{\bibfnamefont{S.~A.} \bibnamefont{Kivelson}},
  \bibnamefont{and} \bibinfo{author}{\bibfnamefont{V.~J.} \bibnamefont{Emery}},
  \bibinfo{journal}{Phys. Rev. B} \textbf{\bibinfo{volume}{62}},
  \bibinfo{pages}{3422} (\bibinfo{year}{2000}).

\bibitem[{\citenamefont{Karakonstantakis
  et~al.}(2011)\citenamefont{Karakonstantakis, Berg, White, and
  Kivelson}}]{Karakonstantakis2011}
\bibinfo{author}{\bibfnamefont{G.}~\bibnamefont{Karakonstantakis}},
  \bibinfo{author}{\bibfnamefont{E.}~\bibnamefont{Berg}},
  \bibinfo{author}{\bibfnamefont{S.~R.} \bibnamefont{White}}, \bibnamefont{and}
  \bibinfo{author}{\bibfnamefont{S.~a.} \bibnamefont{Kivelson}},
  \bibinfo{journal}{Phys. Rev. B} \textbf{\bibinfo{volume}{83}},
  \bibinfo{pages}{054508} (\bibinfo{year}{2011}).

\bibitem[{\citenamefont{White and Scalapino}(2009)}]{White2009}
\bibinfo{author}{\bibfnamefont{S.R.}~\bibnamefont{White}} \bibnamefont{and}
  \bibinfo{author}{\bibfnamefont{D.J.}~\bibnamefont{Scalapino}},
  \bibinfo{journal}{Phys. Rev. B} \textbf{\bibinfo{volume}{79}},
  \bibinfo{pages}{220504(R)} (\bibinfo{year}{2009}).

\bibitem[{\citenamefont{Scalapino and White}(2012)}]{Scalapino2012}
\bibinfo{author}{\bibfnamefont{D.}~\bibnamefont{Scalapino}} \bibnamefont{and}
  \bibinfo{author}{\bibfnamefont{S.}~\bibnamefont{White}},
  \bibinfo{journal}{Physica C Supercond} \textbf{\bibinfo{volume}{481}},
  \bibinfo{pages}{146} (\bibinfo{year}{2012}).

\bibitem[{\citenamefont{Liu et~al.}(2012)\citenamefont{Liu, Yao, Berg, White,
  and Kivelson}}]{Liu2012}
\bibinfo{author}{\bibfnamefont{L.}~\bibnamefont{Liu}},
  \bibinfo{author}{\bibfnamefont{H.}~\bibnamefont{Yao}},
  \bibinfo{author}{\bibfnamefont{E.}~\bibnamefont{Berg}},
  \bibinfo{author}{\bibfnamefont{S.~R.} \bibnamefont{White}}, \bibnamefont{and}
  \bibinfo{author}{\bibfnamefont{S.~A.} \bibnamefont{Kivelson}},
  \bibinfo{journal}{Phys. Rev. Lett.} \textbf{\bibinfo{volume}{108}},
  \bibinfo{pages}{126406} (\bibinfo{year}{2012}).

\bibitem[{\citenamefont{Ehlers et~al.}(2017)\citenamefont{Ehlers, White, and
  Noack}}]{Ehlers2017}
\bibinfo{author}{\bibfnamefont{G.}~\bibnamefont{Ehlers}},
  \bibinfo{author}{\bibfnamefont{S.~R.} \bibnamefont{White}}, \bibnamefont{and}
  \bibinfo{author}{\bibfnamefont{R.~M.} \bibnamefont{Noack}},
  \bibinfo{journal}{Phys. Rev. B} \textbf{\bibinfo{volume}{95}},
  \bibinfo{pages}{125125} (\bibinfo{year}{2017}).

\bibitem[{\citenamefont{Zheng et~al.}(2017)\citenamefont{Zheng, Chung, Corboz,
  Ehlers, Qin, Noack, Shi, White, Zhang, and Chan}}]{Zheng2017}
\bibinfo{author}{\bibfnamefont{B.-X.} \bibnamefont{Zheng}},
  \bibinfo{author}{\bibfnamefont{C.-M.} \bibnamefont{Chung}},
  \bibinfo{author}{\bibfnamefont{P.}~\bibnamefont{Corboz}},
  \bibinfo{author}{\bibfnamefont{G.}~\bibnamefont{Ehlers}},
  \bibinfo{author}{\bibfnamefont{M.-P.} \bibnamefont{Qin}},
  \bibinfo{author}{\bibfnamefont{R.~M.} \bibnamefont{Noack}},
  \bibinfo{author}{\bibfnamefont{H.}~\bibnamefont{Shi}},
  \bibinfo{author}{\bibfnamefont{S.~R.} \bibnamefont{White}},
  \bibinfo{author}{\bibfnamefont{S.}~\bibnamefont{Zhang}}, \bibnamefont{and}
  \bibinfo{author}{\bibfnamefont{G.~K.-L.} \bibnamefont{Chan}},
  \bibinfo{journal}{Science} \textbf{\bibinfo{volume}{358}},
  \bibinfo{pages}{1155} (\bibinfo{year}{2017}).

\bibitem[{\citenamefont{Giamarchi}(2004)}]{Giamarchi2004}
\bibinfo{author}{\bibfnamefont{T.}~\bibnamefont{Giamarchi}},
  \bibinfo{journal}{Chem. Rev.} \textbf{\bibinfo{volume}{104}},
  \bibinfo{pages}{5037} (\bibinfo{year}{2004}).

\bibitem[{\citenamefont{J{\'{e}}rome et~al.}(1989)\citenamefont{J{\'{e}}rome,
  Creuzet, and Bourbonnais}}]{Jerome1989}
\bibinfo{author}{\bibfnamefont{D.}~\bibnamefont{J{\'{e}}rome}},
  \bibinfo{author}{\bibfnamefont{F.}~\bibnamefont{Creuzet}}, \bibnamefont{and}
  \bibinfo{author}{\bibfnamefont{C.}~\bibnamefont{Bourbonnais}},
  \bibinfo{journal}{Physica Scripta} \textbf{\bibinfo{volume}{T27}},
  \bibinfo{pages}{130} (\bibinfo{year}{1989}).

\bibitem[{Note1()}]{Note1}
Note1, \bibinfo{note}{experimentally, this distance is controlled either by
  applying external pressure to the sample, or chemically (choice of anions
  between the cationic stacks of molecules that make up the 1D chains)}.

\bibitem[{\citenamefont{Shinagawa et~al.}(2007)\citenamefont{Shinagawa,
  Kurosaki, Zhang, Parker, Brown, J{\'{e}}rome, Christensen, and
  Bechgaard}}]{Shinagawa2007}
\bibinfo{author}{\bibfnamefont{J.}~\bibnamefont{Shinagawa}},
  \bibinfo{author}{\bibfnamefont{Y.}~\bibnamefont{Kurosaki}},
  \bibinfo{author}{\bibfnamefont{F.}~\bibnamefont{Zhang}},
  \bibinfo{author}{\bibfnamefont{C.}~\bibnamefont{Parker}},
  \bibinfo{author}{\bibfnamefont{S.~E.} \bibnamefont{Brown}},
  \bibinfo{author}{\bibfnamefont{D.}~\bibnamefont{J{\'{e}}rome}},
  \bibinfo{author}{\bibfnamefont{J.~B.} \bibnamefont{Christensen}},
  \bibnamefont{and}
  \bibinfo{author}{\bibfnamefont{K.}~\bibnamefont{Bechgaard}},
  \bibinfo{journal}{Phys. Rev. Lett.} \textbf{\bibinfo{volume}{98}},
  \bibinfo{pages}{147002} (\bibinfo{year}{2007}).

\bibitem[{\citenamefont{Yonezawa et~al.}(2012)\citenamefont{Yonezawa, Maeno,
  Bechgaard, and J{\'{e}}rome}}]{Yonezawa2012}
\bibinfo{author}{\bibfnamefont{S.}~\bibnamefont{Yonezawa}},
  \bibinfo{author}{\bibfnamefont{Y.}~\bibnamefont{Maeno}},
  \bibinfo{author}{\bibfnamefont{K.}~\bibnamefont{Bechgaard}},
  \bibnamefont{and}
  \bibinfo{author}{\bibfnamefont{D.}~\bibnamefont{J{\'{e}}rome}},
  \bibinfo{journal}{Phys. Rev. B} \textbf{\bibinfo{volume}{85}},
  \bibinfo{pages}{140502(R)} (\bibinfo{year}{2012}).

\bibitem[{\citenamefont{Sigrist and Ueda}(1991)}]{Sigrist1991}
\bibinfo{author}{\bibfnamefont{M.}~\bibnamefont{Sigrist}} \bibnamefont{and}
  \bibinfo{author}{\bibfnamefont{K.}~\bibnamefont{Ueda}},
  \bibinfo{journal}{Rev. Mod. Phys.} \textbf{\bibinfo{volume}{63}},
  \bibinfo{pages}{239} (\bibinfo{year}{1991}).

\bibitem[{\citenamefont{Hasegawa}(1996)}]{Hasegawa1996}
\bibinfo{author}{\bibfnamefont{Y.}~\bibnamefont{Hasegawa}},
  \bibinfo{journal}{J. Phys. Soc. Jpn.} \textbf{\bibinfo{volume}{65}},
  \bibinfo{pages}{3131} (\bibinfo{year}{1996}).

\bibitem[{\citenamefont{Moriya and Ueda}(2003)}]{Moriya2003}
\bibinfo{author}{\bibfnamefont{T.}~\bibnamefont{Moriya}} \bibnamefont{and}
  \bibinfo{author}{\bibfnamefont{K.}~\bibnamefont{Ueda}}, \bibinfo{type}{Tech.
  Rep.} (\bibinfo{year}{2003}).

\bibitem[{\citenamefont{Chubukov et~al.}(2008)\citenamefont{Chubukov, Efremov,
  and Eremin}}]{Chubukov2008}
\bibinfo{author}{\bibfnamefont{A.~V.} \bibnamefont{Chubukov}},
  \bibinfo{author}{\bibfnamefont{D.~V.} \bibnamefont{Efremov}},
  \bibnamefont{and} \bibinfo{author}{\bibfnamefont{I.}~\bibnamefont{Eremin}},
  \bibinfo{journal}{Phys. Rev. B} \textbf{\bibinfo{volume}{78}},
  \bibinfo{pages}{134512} (\bibinfo{year}{2008}).

\bibitem[{\citenamefont{Vladimirov et~al.}(2011)\citenamefont{Vladimirov, Ihle,
  and Plakida}}]{Vladimirov2011}
\bibinfo{author}{\bibfnamefont{A.~A.} \bibnamefont{Vladimirov}},
  \bibinfo{author}{\bibfnamefont{D.}~\bibnamefont{Ihle}}, \bibnamefont{and}
  \bibinfo{author}{\bibfnamefont{N.~M.} \bibnamefont{Plakida}},
  \bibinfo{journal}{Phys. Rev. B} \textbf{\bibinfo{volume}{83}},
  \bibinfo{pages}{024411} (\bibinfo{year}{2011}).

\bibitem[{\citenamefont{James et~al.}(2013)\citenamefont{James, Konik, Huang,
  Chen, Rice, and Zhang}}]{James2013}
\bibinfo{author}{\bibfnamefont{A.~J.~A.} \bibnamefont{James}},
  \bibinfo{author}{\bibfnamefont{R.~M.} \bibnamefont{Konik}},
  \bibinfo{author}{\bibfnamefont{K.}~\bibnamefont{Huang}},
  \bibinfo{author}{\bibfnamefont{W.-Q.} \bibnamefont{Chen}},
  \bibinfo{author}{\bibfnamefont{T.~M.} \bibnamefont{Rice}}, \bibnamefont{and}
  \bibinfo{author}{\bibfnamefont{F.~C.} \bibnamefont{Zhang}},
  \bibinfo{journal}{J. Phys.: Conf. Ser} \textbf{\bibinfo{volume}{449}},
  \bibinfo{pages}{12006} (\bibinfo{year}{2013}).

\bibitem[{\citenamefont{Chen et~al.}(2017)\citenamefont{Chen, Leblanc, and
  Gull}}]{Chen2017a}
\bibinfo{author}{\bibfnamefont{X.}~\bibnamefont{Chen}},
  \bibinfo{author}{\bibfnamefont{J.~P.~F.} \bibnamefont{Leblanc}},
  \bibnamefont{and} \bibinfo{author}{\bibfnamefont{E.}~\bibnamefont{Gull}},
  \bibinfo{journal}{Nat. Commun.} \textbf{\bibinfo{volume}{8}},
  \bibinfo{pages}{14986} (\bibinfo{year}{2017}).

\bibitem[{Note2()}]{Note2}
Note2, \bibinfo{note}{$\protect \ensuremath {N_{\protect \rm ch}}$ larger than
  presented in this work can be handled by pDMRG. As a single calculation can
  be spread out across dozens of nodes i.e. thousands of CPU cores, a given
  problem is less limited by the capacity of the computer as such, but by the
  available time on the supercomputer.}

\bibitem[{\citenamefont{Haldane}(1983)}]{Haldane1983a}
\bibinfo{author}{\bibfnamefont{F.~D.~M.} \bibnamefont{Haldane}},
  \bibinfo{journal}{Phys. Rev. Lett.} \textbf{\bibinfo{volume}{50}},
  \bibinfo{pages}{1153} (\bibinfo{year}{1983}).

\bibitem[{\citenamefont{Dagotto and Rice}(1996)}]{Dagotto1996}
\bibinfo{author}{\bibfnamefont{E.}~\bibnamefont{Dagotto}} \bibnamefont{and}
  \bibinfo{author}{\bibfnamefont{T.~M.} \bibnamefont{Rice}},
  \bibinfo{journal}{Science} \textbf{\bibinfo{volume}{271}},
  \bibinfo{pages}{618} (\bibinfo{year}{1996}).

\bibitem[{\citenamefont{Schulz}(1986)}]{Schulz1986}
\bibinfo{author}{\bibfnamefont{H.~J.} \bibnamefont{Schulz}},
  \bibinfo{journal}{Phys. Rev. B} \textbf{\bibinfo{volume}{34}},
  \bibinfo{pages}{6372} (\bibinfo{year}{1986}).

\bibitem[{\citenamefont{Affleck et~al.}(1989)\citenamefont{Affleck, Gepner,
  Schulz, and Ziman}}]{Affleck1989}
\bibinfo{author}{\bibfnamefont{I.}~\bibnamefont{Affleck}},
  \bibinfo{author}{\bibfnamefont{D.}~\bibnamefont{Gepner}},
  \bibinfo{author}{\bibfnamefont{H.~J.} \bibnamefont{Schulz}},
  \bibnamefont{and} \bibinfo{author}{\bibfnamefont{T.}~\bibnamefont{Ziman}},
  \bibinfo{journal}{J. Phys. A: Math. Gen.} \textbf{\bibinfo{volume}{22}},
  \bibinfo{pages}{511} (\bibinfo{year}{1989}).

\bibitem[{\citenamefont{Kantian and Giamarchi}()}]{Kantiana}
\bibinfo{author}{\bibfnamefont{A.}~\bibnamefont{Kantian}} \bibnamefont{and}
  \bibinfo{author}{\bibfnamefont{T.}~\bibnamefont{Giamarchi}},
  \bibinfo{journal}{in preparation}.

\bibitem[{\citenamefont{Ejima et~al.}(2005)\citenamefont{Ejima, Gebhard,
  Nishimoto, and Ohta}}]{Ejima2005}
\bibinfo{author}{\bibfnamefont{S.}~\bibnamefont{Ejima}},
  \bibinfo{author}{\bibfnamefont{F.}~\bibnamefont{Gebhard}},
  \bibinfo{author}{\bibfnamefont{S.}~\bibnamefont{Nishimoto}},
  \bibnamefont{and} \bibinfo{author}{\bibfnamefont{Y.}~\bibnamefont{Ohta}},
  \bibinfo{journal}{Phys. Rev. B} \textbf{\bibinfo{volume}{72}},
  \bibinfo{pages}{033101} (\bibinfo{year}{2005}).

\bibitem[{\citenamefont{Mila and Zotos}(1993)}]{Mila1993}
\bibinfo{author}{\bibfnamefont{F.}~\bibnamefont{Mila}} \bibnamefont{and}
  \bibinfo{author}{\bibfnamefont{X.}~\bibnamefont{Zotos}},
  \bibinfo{journal}{Europhys. Lett.} \textbf{\bibinfo{volume}{24}},
  \bibinfo{pages}{133} (\bibinfo{year}{1993}).

\bibitem[{\citenamefont{Bauer et~al.}()\citenamefont{Bauer, Ewart, Dolfi,
  Kantian, Kosenkov, Keller, and Troyer}}]{Bauera}
\bibinfo{author}{\bibfnamefont{B.}~\bibnamefont{Bauer}},
  \bibinfo{author}{\bibfnamefont{T.}~\bibnamefont{Ewart}},
  \bibinfo{author}{\bibfnamefont{M.}~\bibnamefont{Dolfi}},
  \bibinfo{author}{\bibfnamefont{A.}~\bibnamefont{Kantian}},
  \bibinfo{author}{\bibfnamefont{A.}~\bibnamefont{Kosenkov}},
  \bibinfo{author}{\bibfnamefont{S.}~\bibnamefont{Keller}}, \bibnamefont{and}
  \bibinfo{author}{\bibfnamefont{M.}~\bibnamefont{Troyer}},
  \emph{\bibinfo{title}{https://maquis-ch.github.io/dmrg/}},
  \urlprefix\url{https://maquis-ch.github.io/dmrg/}.

\bibitem[{\citenamefont{White and Chernyshev}(2007)}]{White2007}
\bibinfo{author}{\bibfnamefont{S.R.}~\bibnamefont{White}} \bibnamefont{and}
  \bibinfo{author}{\bibfnamefont{A.L.}~\bibnamefont{Chernyshev}},
  \bibinfo{journal}{Phys. Rev. Lett.} \textbf{\bibinfo{volume}{99}},
  \bibinfo{pages}{127004} (\bibinfo{year}{2007}).

\bibitem[{Note3()}]{Note3}
Note3, \bibinfo{note}{a transition to a CDW would happen in our systems
  eventually if $V/t$ were to be increased further}.

\bibitem[{\citenamefont{Lake et~al.}(1999)\citenamefont{Lake, Aeppli, Mason,
  Schroder, McMorrow, Lefmann, Isshiki, Nohara, Takagi, and Hayden}}]{Lake1999}
\bibinfo{author}{\bibfnamefont{B.}~\bibnamefont{Lake}},
  \bibinfo{author}{\bibfnamefont{G.}~\bibnamefont{Aeppli}},
  \bibinfo{author}{\bibfnamefont{T.~E.} \bibnamefont{Mason}},
  \bibinfo{author}{\bibfnamefont{A.}~\bibnamefont{Schroder}},
  \bibinfo{author}{\bibfnamefont{D.~F.} \bibnamefont{McMorrow}},
  \bibinfo{author}{\bibfnamefont{K.}~\bibnamefont{Lefmann}},
  \bibinfo{author}{\bibfnamefont{M.}~\bibnamefont{Isshiki}},
  \bibinfo{author}{\bibfnamefont{M.}~\bibnamefont{Nohara}},
  \bibinfo{author}{\bibfnamefont{H.}~\bibnamefont{Takagi}}, \bibnamefont{and}
  \bibinfo{author}{\bibfnamefont{S.~M.} \bibnamefont{Hayden}},
  \bibinfo{journal}{Nature} \textbf{\bibinfo{volume}{400}}, \bibinfo{pages}{43}
  (\bibinfo{year}{1999}).

\bibitem[{\citenamefont{Dai et~al.}(2001)\citenamefont{Dai, Mook, Hunt, and
  Doğan}}]{Dai2001}
\bibinfo{author}{\bibfnamefont{P.}~\bibnamefont{Dai}},
  \bibinfo{author}{\bibfnamefont{H.A.}~\bibnamefont{Mook}},
  \bibinfo{author}{\bibfnamefont{R.D.}~\bibnamefont{Hunt}}, \bibnamefont{and}
  \bibinfo{author}{\bibfnamefont{F.}~\bibnamefont{Dogan}},
  \bibinfo{journal}{Phys. Rev. B} \textbf{\bibinfo{volume}{63}},
  \bibinfo{pages}{054525} (\bibinfo{year}{2001}).

\bibitem[{\citenamefont{Razzoli et~al.}(2013)\citenamefont{Razzoli, Drachuck,
  Keren, Radovic, Plumb, Chang, Huang, Ding, Mesot, and Shi}}]{Razzoli2013}
\bibinfo{author}{\bibfnamefont{E.}~\bibnamefont{Razzoli}},
  \bibinfo{author}{\bibfnamefont{G.}~\bibnamefont{Drachuck}},
  \bibinfo{author}{\bibfnamefont{A.}~\bibnamefont{Keren}},
  \bibinfo{author}{\bibfnamefont{M.}~\bibnamefont{Radovic}},
  \bibinfo{author}{\bibfnamefont{N.~C.} \bibnamefont{Plumb}},
  \bibinfo{author}{\bibfnamefont{J.}~\bibnamefont{Chang}},
  \bibinfo{author}{\bibfnamefont{Y.-B.} \bibnamefont{Huang}},
  \bibinfo{author}{\bibfnamefont{H.}~\bibnamefont{Ding}},
  \bibinfo{author}{\bibfnamefont{J.}~\bibnamefont{Mesot}}, \bibnamefont{and}
  \bibinfo{author}{\bibfnamefont{M.}~\bibnamefont{Shi}},
  \bibinfo{journal}{Phys. Rev. Lett.} \textbf{\bibinfo{volume}{110}},
  \bibinfo{pages}{047004} (\bibinfo{year}{2013}).

\bibitem[{\citenamefont{Cho et~al.}(2013)\citenamefont{Cho, Thomale, Raghu, and
  Kivelson}}]{Cho2013}
\bibinfo{author}{\bibfnamefont{W.}~\bibnamefont{Cho}},
  \bibinfo{author}{\bibfnamefont{R.}~\bibnamefont{Thomale}},
  \bibinfo{author}{\bibfnamefont{S.}~\bibnamefont{Raghu}}, \bibnamefont{and}
  \bibinfo{author}{\bibfnamefont{S.~A.} \bibnamefont{Kivelson}},
  \bibinfo{journal}{Phys. Rev. B} \textbf{\bibinfo{volume}{88}},
  \bibinfo{pages}{064505} (\bibinfo{year}{2013}).

\bibitem[{\citenamefont{Albuquerque et~al.}(2007)\citenamefont{Albuquerque,
  Alet, Corboz, Dayal, Feiguin, Fuchs, Gamper, Gull, G{\"{u}}rtler, Honecker
  et~al.}}]{Albuquerque2007}
\bibinfo{author}{\bibfnamefont{A.}~\bibnamefont{Albuquerque}},
  \bibinfo{author}{\bibfnamefont{F.}~\bibnamefont{Alet}},
  \bibinfo{author}{\bibfnamefont{P.}~\bibnamefont{Corboz}},
  \bibinfo{author}{\bibfnamefont{P.}~\bibnamefont{Dayal}},
  \bibinfo{author}{\bibfnamefont{A.}~\bibnamefont{Feiguin}},
  \bibinfo{author}{\bibfnamefont{S.}~\bibnamefont{Fuchs}},
  \bibinfo{author}{\bibfnamefont{L.}~\bibnamefont{Gamper}},
  \bibinfo{author}{\bibfnamefont{E.}~\bibnamefont{Gull}},
  \bibinfo{author}{\bibfnamefont{S.}~\bibnamefont{G{\"{u}}rtler}},
  \bibinfo{author}{\bibfnamefont{A.}~\bibnamefont{Honecker}},
  \bibnamefont{et~al.}, \bibinfo{journal}{J. Magn. Magn. Mater.}
  \textbf{\bibinfo{volume}{310}}, \bibinfo{pages}{1187} (\bibinfo{year}{2007}).

\bibitem[{\citenamefont{Bauer et~al.}(2011)\citenamefont{Bauer, Carr, Evertz,
  Feiguin, Freire, Fuchs, Gamper, Gukelberger, Gull, Guertler
  et~al.}}]{Bauer2011a}
\bibinfo{author}{\bibfnamefont{B.}~\bibnamefont{Bauer}},
  \bibinfo{author}{\bibfnamefont{L.~D.} \bibnamefont{Carr}},
  \bibinfo{author}{\bibfnamefont{H.~G.} \bibnamefont{Evertz}},
  \bibinfo{author}{\bibfnamefont{A.}~\bibnamefont{Feiguin}},
  \bibinfo{author}{\bibfnamefont{J.}~\bibnamefont{Freire}},
  \bibinfo{author}{\bibfnamefont{S.}~\bibnamefont{Fuchs}},
  \bibinfo{author}{\bibfnamefont{L.}~\bibnamefont{Gamper}},
  \bibinfo{author}{\bibfnamefont{J.}~\bibnamefont{Gukelberger}},
  \bibinfo{author}{\bibfnamefont{E.}~\bibnamefont{Gull}},
  \bibinfo{author}{\bibfnamefont{S.}~\bibnamefont{Guertler}},
  \bibnamefont{et~al.}, \bibinfo{journal}{J. Stat. Mech. Theor. Exp.}
  \textbf{\bibinfo{volume}{2011}}, \bibinfo{pages}{P05001}
  (\bibinfo{year}{2011}).

\bibitem[{\citenamefont{Dolfi et~al.}(2014)\citenamefont{Dolfi, Bauer, Keller,
  Kosenkov, Ewart, Kantian, Giamarchi, and Troyer}}]{Dolfi2014a}
\bibinfo{author}{\bibfnamefont{M.}~\bibnamefont{Dolfi}},
  \bibinfo{author}{\bibfnamefont{B.}~\bibnamefont{Bauer}},
  \bibinfo{author}{\bibfnamefont{S.}~\bibnamefont{Keller}},
  \bibinfo{author}{\bibfnamefont{A.}~\bibnamefont{Kosenkov}},
  \bibinfo{author}{\bibfnamefont{T.}~\bibnamefont{Ewart}},
  \bibinfo{author}{\bibfnamefont{A.}~\bibnamefont{Kantian}},
  \bibinfo{author}{\bibfnamefont{T.}~\bibnamefont{Giamarchi}},
  \bibnamefont{and} \bibinfo{author}{\bibfnamefont{M.}~\bibnamefont{Troyer}},
  \bibinfo{journal}{Comput. Phys. Commun.} \textbf{\bibinfo{volume}{185}},
  \bibinfo{pages}{3430} (\bibinfo{year}{2014}).

\bibitem[{\citenamefont{Dolfi et~al.}(2012)\citenamefont{Dolfi, Bauer, Troyer,
  and Ristivojevic}}]{Dolfi2012}
\bibinfo{author}{\bibfnamefont{M.}~\bibnamefont{Dolfi}},
  \bibinfo{author}{\bibfnamefont{B.}~\bibnamefont{Bauer}},
  \bibinfo{author}{\bibfnamefont{M.}~\bibnamefont{Troyer}}, \bibnamefont{and}
  \bibinfo{author}{\bibfnamefont{Z.}~\bibnamefont{Ristivojevic}},
  \bibinfo{journal}{Phys. Rev. Lett.} \textbf{\bibinfo{volume}{109}},
  \bibinfo{pages}{20604} (\bibinfo{year}{2012}).

\bibitem[{\citenamefont{Bauer et~al.}(2012)\citenamefont{Bauer, Corboz,
  L{\"{a}}uchli, Messio, Penc, Troyer, and Mila}}]{Bauer2012}
\bibinfo{author}{\bibfnamefont{B.}~\bibnamefont{Bauer}},
  \bibinfo{author}{\bibfnamefont{P.}~\bibnamefont{Corboz}},
  \bibinfo{author}{\bibfnamefont{A.M.}~\bibnamefont{L{\"{a}}uchli}},
  \bibinfo{author}{\bibfnamefont{L.}~\bibnamefont{Messio}},
  \bibinfo{author}{\bibfnamefont{K.}~\bibnamefont{Penc}},
  \bibinfo{author}{\bibfnamefont{M.}~\bibnamefont{Troyer}}, \bibnamefont{and}
  \bibinfo{author}{\bibfnamefont{F.}~\bibnamefont{Mila}},
  \bibinfo{journal}{Phys. Rev. B} \textbf{\bibinfo{volume}{85}},
  \bibinfo{pages}{125116} (\bibinfo{year}{2012}).

\bibitem[{\citenamefont{Bauer et~al.}(2013)\citenamefont{Bauer, Keller, Dolfi,
  Trebst, and Ludwig}}]{Bauer2013}
\bibinfo{author}{\bibfnamefont{B.}~\bibnamefont{Bauer}},
  \bibinfo{author}{\bibfnamefont{B.~P.} \bibnamefont{Keller}},
  \bibinfo{author}{\bibfnamefont{M.}~\bibnamefont{Dolfi}},
  \bibinfo{author}{\bibfnamefont{S.}~\bibnamefont{Trebst}}, \bibnamefont{and}
  \bibinfo{author}{\bibfnamefont{A.~W.~W.} \bibnamefont{Ludwig}},
  p.~\bibinfo{pages}{5} (\bibinfo{year}{2013}), \eprint{1303.6963}.

\bibitem[{\citenamefont{Bauer et~al.}(2014)\citenamefont{Bauer, Cincio, Keller,
  Dolfi, Vidal, Trebst, and Ludwig}}]{Bauer2014}
\bibinfo{author}{\bibfnamefont{B.}~\bibnamefont{Bauer}},
  \bibinfo{author}{\bibfnamefont{L.}~\bibnamefont{Cincio}},
  \bibinfo{author}{\bibfnamefont{B.}~\bibnamefont{Keller}},
  \bibinfo{author}{\bibfnamefont{M.}~\bibnamefont{Dolfi}},
  \bibinfo{author}{\bibfnamefont{G.}~\bibnamefont{Vidal}},
  \bibinfo{author}{\bibfnamefont{S.}~\bibnamefont{Trebst}}, \bibnamefont{and}
  \bibinfo{author}{\bibfnamefont{A.}~\bibnamefont{Ludwig}},
  \bibinfo{journal}{Nat. Commun.} \textbf{\bibinfo{volume}{5}},
  \bibinfo{pages}{5137} (\bibinfo{year}{2014}).

\bibitem[{\citenamefont{Dolfi et~al.}(2015)\citenamefont{Dolfi, Bauer, Keller,
  and Troyer}}]{Dolfi2015b}
\bibinfo{author}{\bibfnamefont{M.}~\bibnamefont{Dolfi}},
  \bibinfo{author}{\bibfnamefont{B.}~\bibnamefont{Bauer}},
  \bibinfo{author}{\bibfnamefont{S.}~\bibnamefont{Keller}}, \bibnamefont{and}
  \bibinfo{author}{\bibfnamefont{M.}~\bibnamefont{Troyer}},
  \bibinfo{journal}{Phys. Rev. B} \textbf{\bibinfo{volume}{92}},
  \bibinfo{pages}{195139} (\bibinfo{year}{2015}).

\end{thebibliography}

\appendix \label{sec:app}

\section{Spin gap data at finite length and scaling procedure to $\Ltoinf$.}\label{app:scale}
In this section we provide background on the linear scaling procedure we have used in Fig.~\ref{fig:scalgap}
to obtain the infinite-sized gaps summarized in Tab.~\ref{tab:spingaps} and shown in Fig.~\ref{fig:gapNch}. 
We also list all $\Dsl{L,\Nch}$-values that
entered into performing those scalings, in Tab.~\ref{tab:Es_raw}.

There are three related reasons why we perform the fitting of $\Dsl{L,\Nch}$ in Sec.~\ref{subsec:spingap} with 
a linear polynomial in $1/L$: \textbf{(i)}~we find that in the noninteracting regime, $U=V=0$, the averaged spin gap
$\Dsl{L,\Nch}$ shows behaviour congruent with purely linear scaling with weak superimposed oscillations, while a fitting with a quadratic polynomial yields 
physically nonsensical answers. \textbf{(ii)}~mirroring this effect for the interacting systems, we find that
the amplitude of the oscillatory remnant that survives in the averaged spin gap seems to decrease so slowly with $1/L$ that
the oscillations swamp any weak quadratic dependency, should there be any, For this reason, fitting the limited number of available data points with a quadratic
polynomial yields unstable results, while a linear fit works stably.
\textbf{(iii)}~we know by comparing to the existing theory of the $\Nch=2$ ladders that a linear polynomial fit yields physically consistent
results that will, at minimum, provide correct trends for the behaviour of $\Dsl{\Ltoinf,\Nch}$ with $\Nch$.

To take each of the points (i) - (iii) in turn:
%BEGIN FIGURE%
\begin{figure*}
\includegraphics[width=\textwidth]{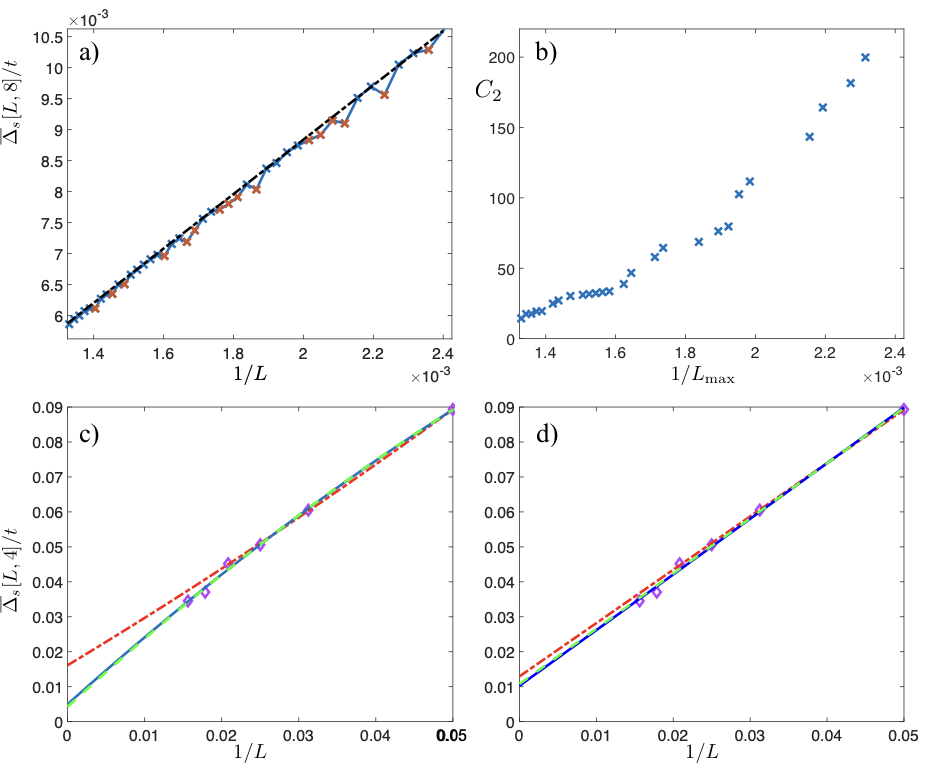}
\caption{
\textbf{(a)} and \textbf{(b)}: decomposing $\Dsl{L,\Nch}$ into linear part and oscillatory remnant, 
and ruling out a quadratic component to the linear part in the noninteracting system (example for $\Nch=8$).
\textbf{(a)}~$\Dsl{L,8}$ at $\tp/t=0.1$ and half-filling, split into a seemingly linear component (blue crosses, linear fit as dash-dotted line)
and weak oscillatory remnant (red crosses).
\textbf{(b)}~Confirming that the seemingly linear component in (a) is actually linear by plotting the $1/L^2$-coefficient of a second-order 
polynomial fitting as a function of the upper fitting range $L_{\rm max}$ (see text for discussion).
\textbf{(c)} and \textbf{(d)}: Instability and stability of fitting procedures respectively for the interacting systems, for example of 
$\Dsl{L,4}$ at $U/t=4$, $V/t=1$, $\tp/t=0.1$ (magenta diamonds in both figures).
\textbf{(c)} Fitting with second-order polynomial for four largest $\Dsl{L,4}$-values (red dash-dotted line), five largest values (green dashed line), and all six values (blue line).
\textbf{(d)} Fitting with first-order polynomial, line styles match the same fitting ranges as in (c).
}
\label{fig:scalingjustification1}
\end{figure*}
%END FIGURE%
\textbf{(i):} Consider the noninteracting version of our model Hamiltonian, $U=V=0$, at half-filling and $\tp/t=0.1$.
To take $\Nch=8$ as an arbitrary example, analysing the averaged spin gap $\Dsl{L,8}$ as a function of $1/L$ reveals an
universal pattern in these tunneling-anisotropic systems: while much reduced over
the non-averaged gap $\dsl{L,8}$, weak oscillations persist in $\overline{\Delta}_s$, as shown for this example in Fig.~\ref{fig:scalingjustification1}a.
We find this data to be naturally decomposing into a subset of points obeying perfect linear scaling (blue crosses in 
Fig.~\ref{fig:scalingjustification1}a, linear fit superimposed as dash-dotted line), and the complementary subset of points that is affected by the downwards
oscillatory deviations from this scaling (red crosses in Fig.~\ref{fig:scalingjustification1}a). That the seemingly linear subset of points is actually a linear one 
becomes apparent if one tries to fit it with a second order polynomial $C_0+C_1/L+C_2/L^2$. In Fig.~\ref{fig:scalingjustification1}b we plot the resulting
$C_2$ as a function of the maximal upper fitting range, $L_{\rm max}$. It is apparent that $C_2$ cannot be any underlying physical property
of the system like e.g. a remnant of a band-curvature that survives the averaging, in which case it should converge towards a constant value with increasing $L_{\rm max}$.
Rather, $C_2$ continues to drop even at the large $L_{\rm max}$ shown here, because that is the only way a second-order polynomial fit can 
maintain optimal approximation to a purely linear distribution of data points as its fitting range grows. If any dependence on $1/L^2$ remains within $\Dsl{L,\Nch}$ in the noninteracting regime, the oscillatory remnant clearly dominates it completely.

\textbf{(ii):} Just as for the noninteracting system, remnants of oscillatory behaviour remain visible as we compute
$\Dsl{L,\Nch}$ at $U/t=4$, $V/t=0.5,1$. But unlike the noninteracting regime, each data point now comes with substantial computational cost.
The limited number of available data points compared to the non-interacting regime means that they cannot be reliably divided into a purely linear
subset and an oscillatory remnant. As a consequence, the impact of the weak oscillatory remnant on any scaling $\Ltoinf$ must be considered, especially as
the amplitude of oscillations appears to decay only weakly with $1/L$, as in the non-interacting case (i). 
In Figs.~\ref{fig:scalingjustification1}c and d we contrast the instability of any second-order polynomial fit depending on the fitting range with the respective stability
of a linear fit over the same fitting ranges. This illustrates again, in a way different from (i), how the oscillatory remnant
overwhelms any quadratic component in $1/L$, if such were to exist in the interacting regime (as opposed to the non-interacting one, where we can rule out its existence within the linear subset).

\textbf{(iii):} Complementing the argument of (i) and (ii) so far, that $\Dsl{L,\Nch}$ is dominated by linear behaviour in $1/L$ with weak oscillatory remnants 
that overshadow a potential quadratic contribution, is the physical consistency of linear extrapolation.
For this, we can turn to the $\Nch=2$ case, the 2-leg U-V ladder. The bosonized field theory that describes this ladder at low energy predicts
that both its spin sectors are gapped~\cite{BookGiamarchi2003}. It also shows that both spin gaps must grow with both $U/t$ and $V/t$ (eqs. (8.23), (8.24) and (8.26)
in ref.~\cite{BookGiamarchi2003}). This effect is clearly recovered by our linear extrapolation, as shown in Tab.~\ref{tab:spingaps}. This further supports
that a linear extrapolation, even one that necessarily averages over the weak oscillatory remnant, correctly captures the spin gaps behaviour in the thermodynamic limit
and in particular its change with the systems' parameters. 

%BEGIN TABLE%
\begin{table}[t]
\scriptsize%\subfloat{
    \begin{tabular}{ c || c | c | c | c || c | c | c |}
    
                            &  \multicolumn{4}{|c|}{ $V/t=0.5$ }  &  \multicolumn{3}{|c|}{ $V/t=1$ }  \\ \cline{2-8}
    $L$   & $\Nch=2$ & $\Nch=4$ & $\Nch=6$ & $\Nch=8$ & $\Nch=2$ & $\Nch=4$ & $\Nch=6$\\ \hline\hline
    $20$ & $95.953$ & $96.945$ & $96.956$ & $97.023$   & $87.977$ & $89.311$ & $89.516$ \\ \hline
    $32$ & $66.272$ & $65.597$ & $64.519$ & $64.799$   & $61.726$ & $60.503$ & $61.011$ \\ \hline
    $40$ & $53.981$ & $53.943$ & $54.949$ & $52.908$   & $51.298$ & $50.607$ & $51.774$ \\ \hline
    $48$ & $45.559$ & $46.421$ & $45.634$ & $       $       & $43.520$ & $45.186$ & $42.838$ \\ \hline
    $56$ & $39.519$ & $41.736$ & $39.809$ & $       $       & $37.810$ & $37.049$ & $           $ \\ \hline
    $60$ & $37.113$ & $           $ & $           $ & $           $   & $35.515$ & $           $ & $           $         \\ \hline
    $64$ & $35.024$ & $35.634$ & $           $ & $           $   & $33.508$ & $34.508$ & $           $           \\ \hline
    $72$ & $31.646$ & $           $ & $           $ & $           $   & $30.193$ & $           $ & $           $          \\ \hline
    $80$ & $29.213$ & $           $ & $           $ & $           $   & $27.619$ & $           $ & $           $          \\ \hline
    $96$ & $24.326$ & $           $ & $           $ & $           $   & $24.061$ & $           $ & $           $          \\ \hline
    $120$ & $19.618$ & $           $ & $           $ & $           $ & $19.231$ & $           $ & $           $          \\ \hline
    $160$ & $14.991$ & $           $ & $           $ & $           $ & $           $ & $           $ & $           $              \\ \hline

    \end{tabular}
    
    %}
\caption{Spin gaps $\Dsl{L,\Nch}$ (in units of $t\times 10^{-3}$) upon which extrapolations $\Ltoinf$ were performed in Figs.~\ref{fig:scalgap}
and for Fig.~\ref{fig:altscalings}.}
\label{tab:Es_raw}
\end{table}

%%%%%%%%%%%%%%%%%%%%%%%%%%%%%%%%%%%%%%%%%%%%%%%%%%
%%%%%%%%%%%%%%%%%%%%%%%%%%%%%%%%%%%%%%%%%%%%%%%%%%

\section{Handling large $\chi$ and large MPO bond dimension simultaneously through pDMRG}\label{app:dmrg_tech}

%BEGIN FIGURE%
\begin{figure}[h]
\centering
\includegraphics[width=1\columnwidth]{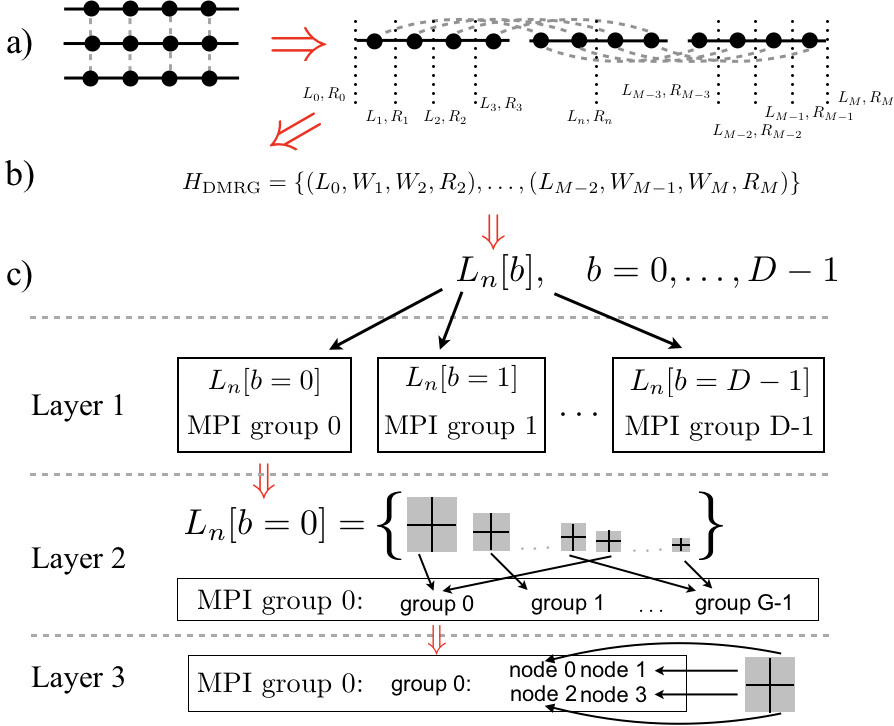}
\caption{
Overview of pDMRG:
\textbf{(a)} The physical system, e.g. a $M$-site 2D lattice, is mapped to the 1D DMRG-chain. Every one of the possible $M+1$ bipartitions, carries left and right boundaries $L_n$, $R_n$ (c.f. Sec.~\ref{sec:method}), typically by far the largest objects in any simulation that has sizeable long-range terms in the Hamiltonian.
\textbf{(b)} For every bipartition, the Hamiltonian consists of the (small) local MPOs $W_n$, $W_{n+1}$ on sites to the left/right of the bipartition, together with the boundaries $L_{n-1}$, $R_{n+1}$.
\textbf{(c)} The three layers of parallelization our pDMRG offers.
In Layer 1, each element $L_n[b]$ of $L_n$ is dispatched to a MPI-group (each of which may be split into $G$ subgroups). Every $L[b_l]$ is a block-sparse matrix. In Layer 2, every one of the dense matrices making up $L[b]$ is distributed round-robin over the the $G$ split-off subgroups held by a global MPI-group.
In Layer 3, the tiles making up a dense block can are distributed among the physical compute nodes held by each split subgroup.
}
\label{fig:pdmrg_full_scheme}
\end{figure}
%END FIGURE%
To make DMRG capable of handling simulations that would exhaust the RAM of any single compute node, and in order to bring the many CPU-cores of modern parallel supercomputers to bear on such large-scale problems, the groups of T. Giamarchi (U. Geneva) and M. Troyer (ETHZ)  have developed pDMRG~\cite{Bauera}, with financial support from the Swiss government under its' HP2C initiative, with a focus on easy extendability and optimizability. Key features are (i) a clean tensor network framework for development of new algorithms; (ii) use of a matrix-like data storage type for direct use of BLAS/LAPACK routines; (iii) implementation of  an arbitrary number of Abelian symmetries, reducing the matrix complexity by use of block-sparse matrices in models with conserved quantum numbers; (iv) bindings to the ALPS library~\cite{Albuquerque2007,Bauer2011a} for a generic model description. Parallelization uses the Message Passing Interface (MPI) standard for distributed memory and Intel\textsuperscript{\textregistered} Cilk{\texttrademark} Plus tasks for multi-threading. The shared memory version of the code has been already published~\cite{Dolfi2014a}, and is used extensively in 1D and 2D condensed matter physics ~\cite{Dolfi2012, Bauer2012, Bauer2013, Bauer2014,Dolfi2015b}. The resulting parallelism of pDMRG is fully scalable and can be run on anything from a single CPU core to hundreds of compute nodes.

The stages of pDMRG have been summarized in Fig.~\ref{fig:pdmrg_full_scheme}. As in regular DMRG, the physical problem is exactly mapped onto a 1D chain. In any realistic problem requiring the use of pDMRG, there will be many long-range interaction and/or tunneling terms along the chain. As a result, the left-/right-boundary pairs $L_n$, $R_n$ at the $n=0,\dots,M$ bipartitions of the DMRG-chain will become objects requiring very large amounts of memory for any problem with a substantial bond dimension $\chi$ (Fig.~\ref{fig:pdmrg_full_scheme}a).
For every bipartition of the DMRG-chain the DMRG-Hamiltonian will be a set of one left- and one right-boundary, together with the (usually small) local MPOs $W_n$, $W_{n+1}$ for the sites $n$ and $n+1$ to the left/right of the $n$-th bipartition (c.f. Fig.~\ref{fig:pdmrg_full_scheme}b). The technical challenge for any DMRG-implementation aiming to parallelize the repeated application of any of these Hamiltonians to a tensor-decomposed wave-function, which is the linear-algebra operation at the heart of any ground-state DMRG, is to distribute the boundaries across the nodes of a parallel supercomputer. Our pDMRG implementation offers three layers for this, summarized in Fig.~\ref{fig:pdmrg_full_scheme}c, making extensive use of the Message Passing Interface (MPI), the de-facto standard for parallel supercomputing:

\textbf{(i)} Layer 1 exploits that the boundaries are effectively vectors of block-sparse matrices (c.f. eq.~(\ref{eq:mpo}) and following, and Fig. ~\ref{fig:pdmrg}c). It distributes the $D$ elements of both these vectors, each element being a block-sparse matrix, in a size-ordered, round-robin fashion within the global group of MPI-groups. Each of the MPI-groups within the global group may encompass several nodes of the compute cluster.

\textbf{(ii)} Inside each MPI-group, Layer 2 then may further distribute the dense matrices comprising each boundary element among MPI-subgroups into which the main group can be split.

\textbf{(iii)} As dense matrices can themselves become large, they are always stored in tiled format. For Layer 3, these tiles may be distributed to different compute nodes within a subgroup.

In the Jacobi-Davidson eigensolver of pDMRG, the tensor contractions become dot-products of vectors of matrices. For this, MPI groups communicate via \verb|MPI_Allreduce|. When shifting to the next pair of sites, the number of boundaries can change, which results in a redistribution of elements among the MPI-processes, using asynchronous point-to-point transfers. On each node, operations are executed in two steps. A dry-run collects all linear algebra operations (mostly \verb|GEMM| and \verb|AXPY|). In the second step the directed acyclic operation-graph is executed in parallel on several threads. This abstract approach optimizes performance, autonomously exploiting hidden operational parallelism. The basic linear algebra operations are then explicitly unrolled between tiles of the dense matrix and dispatched to the underlying BLAS library.
%%%%%%%%%%%%%%%%%%%%%%%%%%%%%%%%%%%%%%%%%%%%%%%%%%

\end{document}